\documentclass[]{emulateapj}

\usepackage{etoolbox}
\newbool{SUBMIT}

\boolfalse{SUBMIT}

\usepackage{color}

\pdfpagewidth=\paperwidth 
\pdfpageheight=\paperheight

\newcommand{\sub}[2]{\ensuremath{#1_{\mathrm{#2}}}}
\newcommand{\super}[2]{\ensuremath{#1^{\mathrm{#2}}}}

\newcommand{\unit}[2]{\ensuremath{\textrm{#1}^{#2}}}
\newcommand{\half}{\frac{1}{2}}
\newcommand{\vect}[1]{\mathbf{#1}}
\newcommand{\Msun}{\ensuremath{M_{\odot}}}

\ifbool{SUBMIT}{}{

}

\usepackage{graphicx} 
\usepackage{epstopdf} 
\usepackage{amsmath}
\usepackage{hyperref}
\usepackage{natbib}
\definecolor{hcitecolor}{RGB}{0,127,0}
\hypersetup{%
  citecolor=hcitecolor,%
  linkcolor=hcitecolor,%
  urlcolor=hcitecolor%
}%

\shorttitle{\textsc{action-space clustering of tidal streams}}
\shortauthors{\textsc{sanderson, helmi, \& hogg 2014}}
\begin{document}

\title{Action-space clustering of tidal streams to infer the Galactic potential}
\author{Robyn E. Sanderson \& Amina Helmi}
\affil{Kapteyn Astronomical Institute, P.O. Box 800, 9700 AV Groningen, The Netherlands}
\email{sanderson@astro.rug.nl}
\and
\author{David W. Hogg} 
\affil{Center for Cosmology and Particle Physics, Department of Physics, New York University, \\ 4 Washington Place, New York, NY  10003, USA;}
\affil{Max-Planck-Institut f\"ur Astronomie, K\"onigstuhl 17, D-69117 Heidelberg, Germany}

\begin{abstract}
We present a new method for constraining the Milky Way halo gravitational potential by simultaneously fitting multiple tidal streams. This method requires full three-dimensional positions and velocities for all stars in the streams, but does not require identification of any specific stream, nor determination of stream membership for any star. We exploit the principle that the action distribution of stream stars is most clustered---that is, most informative---when the potential used to calculate the actions is closest to the true potential. We measure the amount of clustering with the Kullback-Leibler Divergence (KLD) or relative entropy, a statistical measure of information which also provides conditional uncertainties for our parameter estimates. We show, for toy Gaia-like data in a spherical isochrone potential, that maximizing the KLD of the action distribution relative to a smoother distribution recovers the true values of the potential parameters. The precision depends on the observational errors and the number and type of streams in the sample; we find that with the phase-space structure and observational uncertainties expected in the Gaia red-giant-star data set, we measure the enclosed mass at the average radius of the sample stars accurate to 3\% and precise to 20-40\%. Recovery of the scale radius is precise to 25\%, and is biased 50\% high by the small galactocentric distance range of stars in our mock sample (1-25 kpc, or about three scale radii, with mean 6.5 kpc). About 15 streams, with at least 100 stars per stream, are needed to obtain both upper and lower bounds on the enclosed mass and scale radius when observational errors are taken into account; 20-25 streams are required to stabilize the size of the confidence interval. If radial velocities are provided for stars out to 100 kpc (10 scale radii), the bias in the scale radius measurement is eliminated and all parameters can be determined with $\sim$10\% accuracy and 20\% precision (1.3\% accuracy in the case of the enclosed mass). This finding underlines the need for ground-based spectroscopic follow-up to complete the radial velocity catalog for faint halo stars ($V>17$) observed by Gaia.
  
  \end{abstract}


\section{Introduction}
\label{sec:intro}

Observational evidence in the Milky Way \citep{1999Natur.402...53H,2002ApJ...569..245N,2006ApJ...642L.137B}, other $L_*$ spiral galaxies \citep{2009Natur.461...66M,1538-3881-140-4-962} and cosmological simulations \citep{Helmi2011} indicate that the process of hierarchical accretion has left its mark on our Galaxy in the form of tidal streams.
Not only are these streams spectacular evidence of our Galaxy's tumultuous history, they also provide a unique opportunity to determine the shape and size of its gravitational potential.
At large galactocentric distances where tidal streams are most easily detected, the potential is thought to be dominated by the Galactic dark matter halo, about which very little is currently known.
Yet the means to determine its shape is available in the halo itself: observations show that the stellar halo is indeed made of many dynamically and chemically distinct components \citep{1999Natur.402...53H,0004-637X-698-1-567,0004-637X-738-1-79} that are probably accreted streams. 

Fitting of the neighboring orbits of the stars in a stream, which move as test particles in the Galactic potential, can constrain the Galaxy's mass and shape \citep{2001ApJ...551..294I,2004ApJ...610L..97H,2005ApJ...619..800J,2008MNRAS.389.1391S,2009ApJ...697..207W,Eyre2009,2010ApJ...711...32N,2010ApJ...712..260K,2010ApJ...714..229L,2013ApJ...773L...4V}.
Currently, attempts to do this have been restricted to single streams by the difficulties of obtaining positions and velocities of stream stars, confirming the membership of stars in streams, and exploring high-dimensional parameter spaces  \citep[however, see][]{2010PhDT.......194W}.
In the coming decade, however, the Gaia mission \citep{2001A&A...369..339P} will measure the six-dimensional phase space positions of 150 million stars to unprecedented accuracy, including the largest sample of halo stars in history.
Included in this revolutionary data set will be hundreds of streams, the fossil record of the Milky Way's accretion history \citep{1999MNRAS.307..495H,Helmi2002,Gomez2010}.
All this new data will enable---and require---a transition from individual stream-fitting to simultaneous treatment of many streams whose membership is not always certain.

One way to analyze many streams simultaneously is to work in the space of integrals of motion or actions, where the distribution is much simpler than in the space of positions and velocities.
In particular, for integrable potentials the actions of stream stars are adiabatic invariants, so stream stars will be tightly clustered in actions because they originate from small satellite galaxies that initially occupied small volumes in phase space.
The main qualities of a stream, except for the orbital phase information, can be described simply as a clump in three-dimensional action space \citep{1999MNRAS.307..495H}.
At first glance this may not seem much of an advantage for orbit fitting since the actions depend on the gravitational potential, but as we show in this work, we can exploit this dependence to fit the potential.
As one would intuitively expect, the best choice for the potential is the one where the streams are most tightly clustered in action space, for the same reason that one expects a stream integrated backwards in time to re-form into a galaxy in the correct potential \citep[e.g,][]{1999ApJ...512L.109J,Price-Whelan2013}.
Tuning the potential to ``focus" the streams into the tightest action-space clumps is equivalent to simultaneously fitting a common set of orbits to all the stars in each stream and a common potential to all the streams in a data set
(this implicitly prefers potentials in which streams follow single orbits as closely as possible, which is incorrect and could create a bias; we will return to this at the end of the paper).
Even better---in our formulation of this approach---as long as many of the stars belong to one stream or another it is not necessary to determine the membership of each stream.
This sidesteps one major challenge for the analysis of Gaia's huge data set.

In this work we demonstrate the utility of action-space clustering
  to simultaneously fit multiple tidal streams with a common potential.
To measure the degree of clustering we use the Kullback-Liebler divergence (KLD) or ``relative entropy,''
  described in Section \ref{sec:kld}.
The KLD is a measure of the difference between two distribution functions,
  here we use it to measure the difference between the distribution in actions
  and a ``shuffled'' version of the distribution that corresponds to the product of the marginal one-dimensional distributions.
There are other techniques in the literature that exploit phase-space clustering to fit the potential \citep{2009ApJ...703.1061S,2012MNRAS.419.1951V,2012ApJ...760....2P,2013MNRAS.433.1826S},
  but the KLD has the great advantage that it has an interpretation in terms of probabilistic inference;
  it is related to a difference in log likelihood between two distributions.
This provides us with a method for using the KLD to produce justifiable uncertainties on parameter inferences made with it.
The KLD also has the advantage that it is not particularly tuned to any \emph{form} for the clustering;
  it is just looking for differences in information content or predictive value.
That is, our method does not require compact clusters in phase space,
  just structure;
 and so can be applied not just to streams but also to shells,
  catastrophes, and even possibly hotter kinematic components.

In our tests of the method, we use a mock stellar halo built entirely through accretion, by creating a population of progenitor ``galaxies'' consistent with the luminosity function of the known Milky Way satellite galaxies, radially distributed according to the measured stellar halo density profile, with an orbital-eccentricity distribution consistent with cosmological simulations (Section \ref{subsec:ics}). We integrate the stars of these progenitors as test particles for a range of orbital times in an isochrone potential, which has analytic actions (Section \ref{subsec:pot}). The number of progenitors is chosen to create a mock Milky Way halo with the expected number of thin streams.

We convolve the present-day positions of the ``stars" in the mock halo using the Gaia error model, assuming they are all RGB stars with $M_V=1$ (Section \ref{subsec:conv}), select stars with acceptable parallax and radial-velocity errors, and use a rough energy cutoff to select a subset for clustering analysis (Section \ref{subsec:streamsel}).
We calculate the KLD for a grid of values in parameter space for this subset, with and without Gaia errors, and identify the maximum KLD---the best-fit potential parameters---and the confidence intervals (Section \ref{subsec:klcomp}).
Maps of the KLD surface on the parameter-space grid, presented in Sections \ref{subsec:step1results} and \ref{subsec:step2results}, show that our method recovers the input potential parameters within the confidence interval. In Section \ref{sec:varyNstream} we discuss how the number of streams in the sample affects parameter recovery.
Implications and future work are discussed in Section \ref{sec:concl}.

\section{The Kullback-Leibler divergence}
\label{sec:kld}

The Kullback-Leibler divergence \citep{Kullback1951} is a statistic that compares two different probability distributions. For a continuous random variable $\vect{x}$, the KLD \emph{from} $p(\vect{x})$ \emph{to} $q(\vect{x})$ is defined as
\begin{equation}
\sub{D}{KL}(p : q) \equiv \int p(\vect{x}) \log \frac{p(\vect{x})}{q(\vect{x})} d\vect{x}.\footnote{The logarithm can be any base; for this work we use the natural (base $e$) log, for which the units of the KLD are known as ``nats".  }
\end{equation}
In the case that one has a set of $N_*$ points $\vect{x}_i$
  drawn from $p(\vect{x})$,
  the KLD can be estimated by the sampling approximation
\begin{equation}
  \tilde{D}_{KL}(p : q) = \frac{1}{N_*} \sum_{i}^{N_*} \log \frac{p(\vect{x}_i)}{q(\vect{x}_i)},
\end{equation}
illustrating that the $p(\vect{x})$ in the integral form is acting as the measure, $\mu(\vect{x})$, for the integration, such that $d\mu(\vect{x}) \equiv p(\vect{x}) d\vect{x}$: 
\begin{equation}
\sub{D}{KL}(p : q) \equiv \int \log \frac{p(\vect{x})}{q(\vect{x})} d\mu(\vect{x}).
\end{equation}
The KLD is not symmetric in $p$ and $q$, but is always positive. The more $p$ differs from $q$, the higher the value of $\sub{D}{KL}$; if the two distributions are identical, the KLD is zero. 

We use the KLD to measure the degree of clustering in a multidimensional probability distribution $p$ by comparing it to the product of its marginal distributions $\super{p}{shuf}$. If $p$ is a bivariate Gaussian, for example, then
\begin{equation}
\sub{D}{KL}(p: \super{p}{shuf}) = -\half \log(1- \rho^2),
\end{equation}
where $\rho$ is the correlation coefficient \citep[][Equation 6.12]{kullback1959}. The KLD in this case is independent of the dispersions in each direction. This interpretation of the KLD, which measures clustering independent of range, is also known as the mutual information. 

The KLD is a promising tool for using stellar streams to constrain the Galactic potential because stream stars should be clustered in action space. The clustering of streams in action space is tightest when the correct potential (or the closest one to the real potential) is used to compute the actions from the phase space coordinates $\vect{w}\equiv(\vect{x},\vect{v})$. We represent the Galactic potential by a set of characteristic parameters $\vect{a}$, and consider the probability density $f$ of the stars' actions $\vect{J}(\vect{w};\vect{a})$, which we will denote as
\begin{equation}
f_{\vect{a}}(\vect{J})
\end{equation}
since the shape of the probability density is determined by the choice of Galactic parameters $\vect{a}$. Given this notation, \emph{maximizing} the quantity
\begin{equation}
\label{eq:dkl1}
 \sub{\super{D}{I}}{KL} =\int f_{\vect{a}}\left(\vect{J}\right) \log \frac{f_{\vect{a}}\left(\vect{J}\right)}{\super{f_{\vect{a}}}{shuf}\left(\vect{J}\right)}\ d^3\vect{J},
\end{equation}
subject to the constraint that the functions $f_{\vect{a}}$ and $\super{f_{\vect{a}}}{shuf}$ are properly normalized,
will identify the best-fit parameters, labeled $\vect{a}_0$, as the ones that produce the most clustered distribution of the stars' actions $\vect{J}$. 

The KLD also allows us to estimate the error on $\vect{a}_0$ through its interpretation as the expectation value of the difference in log-likelihood (or posterior probability). Given some data $\vect{x}$, the KLD between two probability distributions $p(\vect{x})$ and $q(\vect{x})$
  is related to the ratio of posterior probabilities, $\mathcal{P}(\mathcal{H}_p|\vect{x})/\mathcal{P}(\mathcal{H}_q|\vect{x})$, of the two hypotheses, $\mathcal{H}_p$ and $\mathcal{H}_q$,
  that $\vect{x}$ are drawn from $p$ or $q$ by \citep[][p.5, eq.2.5]{kullback1959}:
\begin{equation}
\label{eq:kld2post}
\sub{D}{KL}(p : q) = \int \log \frac{\mathcal{P}(\mathcal{H}_p|\vect{x})}{\mathcal{P}(\mathcal{H}_q|\vect{x})} p(\vect{x}) d\vect{x} -\log \frac{\mathcal{P}(\mathcal{H}_p)}{\mathcal{P}(\mathcal{H}_q)},
\end{equation}
where  $\mathcal{P}(\mathcal{H})$ are the prior probabilities of the two hypotheses. Bayes's theorem,
\begin{equation}
 \mathcal{L}(\vect{x}|\mathcal{H})\mathcal{P}(\mathcal{H}) = \mathcal{P}(\mathcal{H}|\vect{x}),
\end{equation}
relates the KLD to the difference in log-likelihood  $\mathcal{L}(\vect{x}|\mathcal{H})$ between the two hypotheses:
\begin{equation}
\label{eq:kld2lik}
\sub{D}{KL}(p : q) = \int \log \frac{\mathcal{L}(\mathcal{H}_p|\vect{x})}{\mathcal{L}(\mathcal{H}_q|\vect{x})} p(\vect{x}) d\vect{x}.
\end{equation}
If we assume a flat prior, so that $ \mathcal{P}(\mathcal{H}_p)=\mathcal{P}(\mathcal{H}_q)$, then the second term of \eqref{eq:kld2post} vanishes and the KLD is directly proportional to the expectation value of the difference in the log of the posterior probability:
\begin{eqnarray}
\label{eq:kld2flatprior}
\sub{D}{KL}(p : q) &=& \int \log \frac{\mathcal{P}(\mathcal{H}_p|\vect{x})}{\mathcal{P}(\mathcal{H}_q|\vect{x})} p(\vect{x}) d\vect{x} = \Big\langle \log \frac{\mathcal{P}(\mathcal{H}_p|\vect{x})}{\mathcal{P}(\mathcal{H}_q|\vect{x})} \Big\rangle_p, \nonumber\\
&&\textrm{if } \mathcal{P}(\mathcal{H}_p)=\mathcal{P}(\mathcal{H}_q).
\end{eqnarray}
One can also break up the logarithm of the ratio of the distributions into a difference of logarithms: 
\begin{equation}
\sub{D}{KL}(p : q) = \int p(\vect{x}) \log p(\vect{x}) - \int p(\vect{x}) \log q(\vect{x}).
\end{equation}
The first term is recognizable as the entropy of the distribution $p$; in this form one can think of the KLD as measuring the amount of information we lose by describing the data $\vect{x}$, really drawn from $p$, with $q$ instead. In terms of the likelihoods of the two distributions, we have
\begin{equation}
 \sub{D}{KL}(p : q) = \langle \log \mathcal{L}(\mathcal{H}_p|\vect{x}) \rangle_p - \langle \log \mathcal{L}(\mathcal{H}_q|\vect{x}) \rangle_p.
\end{equation}
If we are comparing many different $q$ to a single $p$, as we will do for our case, the first term is a constant that normalizes the expectation value of the likelihood of $\mathcal{H}_q$ (the likelihood that $\vect{x}$ are drawn from $q$). This means that in Bayesian theory, this first term must be related to the \emph{evidence}.

In our case, we want to derive a confidence interval around the best-fit parameters by testing how much the clustering of the $\vect{J}$ changes as we vary the potential parameters. Therefore, we want to compare the distribution of $\vect{J}$ produced using the best-fit parameters $\vect{a}_{0}$ with the $\vect{J}$-distribution produced by some other parameters $\sub{\vect{a}}{trial}$. Then we can use the interpretation above to relate the KLD between these distributions to the expectation of their relative probability. Essentially, this answers the question ``how far away from the best-fit parameters do I have to go before the $\vect{J}$-distribution significantly differs from the one I got with the best-fit parameters?''

In formal terms, this means we take distribution $p$ in Equations (\ref{eq:kld2post}-\ref{eq:kld2flatprior}) to be the distribution of the actions for the best-fit parameters
  (found by maximizing $\sub{\super{D}{I}}{KL}$):
\begin{equation}
 p(\vect{x}) \to f_{\vect{a}_0}\left(\vect{J}\right).
\end{equation}
The hypothesis $\mathcal{H}_p$, that the $\vect{x}$ for which the KLD is evaluated are drawn from the distribution $p(x)$, is in our case the hypothesis that the stars' actions $\vect{J}$ are drawn from the distribution corresponding to the best-fit potential parameters $\vect{a}_0$. Since the potential parameters indirectly specify the form of the distribution by determining the actions from the phase-space positions $\vect{w}$, we label $\mathcal{H}_p \to \mathcal{H}_{\vect{a}_0}$.

Equivalently, we take for $q$ the distribution of actions with some other set of parameters $\sub{\vect{a}}{trial}$:
\begin{equation}
 q(\vect{x}) \to f_{\sub{\vect{a}}{trial}}\left(\vect{J}\right).
\end{equation}
The corresponding hypothesis, $\mathcal{H}_q \to \mathcal{H}_{\sub{\vect{a}}{trial}}$, is that the potential parameters $\sub{\vect{a}}{trial}$ produce the distribution that best describes the $\vect{J}$.

Making these substitutions, we find that the KLD
\begin{equation}
\label{eq:dkl2}
\sub{\super{D}{II}}{KL} = \int f_{\sub{\vect{a}}{0}}\left(\vect{J}\right)  \log \frac{f_{\sub{\vect{a}}{0}}\left(\vect{J}\right)}{f_{\sub{\vect{a}}{trial}}\left(\vect{J}\right)}\ d^3\vect{J}
\end{equation}
is therefore related to the expectation value of the difference in log posterior probability of potential parameters $\vect{a}_0$ or $\sub{\vect{a}}{trial}$:
\begin{equation}
\label{eq:kld2interp}
\sub{\super{D}{II}}{KL} = \int \log \frac{\mathcal{P}(\mathcal{H}_{\vect{a}_0} | \vect{J})}{\mathcal{P}(\mathcal{H}_{\sub{\vect{a}}{trial}} | \vect{J})} d\mu\left(\vect{J}(\vect{a}_0)\right) = \Big\langle \log \frac{ \mathcal{P}( \mathcal{H}_{\vect{a}_0} | \vect{J} ) }{ \mathcal{P}(\mathcal{H}_{\sub{\vect{a}}{trial}}|\vect{J}) }\Big\rangle_{\vect{J}}.
\end{equation}
In other words, $\sub{\super{D}{II}}{KL} = 1$ for a given $\sub{\vect{a}}{trial}$ means that the expected value of the log probability ratio, for the distribution of $\vect{J}(\vect{a}_0)$, is equal to 1. Therefore we expect that parameters $\sub{\vect{a}}{trial}$ are $e$ times less likely\footnote{assuming we are using natural logarithms in the KLD definition} than the best-fit $\vect{a}_0$ to produce a $\vect{J}$ distribution consistent with the one produced by using $\vect{a}_0$. Appendix \ref{appx:KLDIllustration} gives a graphical illustration of the method using a simple example.

In practice, we first determine the best-fit $\vect{a}_0$ by maximizing $\sub{\super{D}{I}}{KL}$. Then we calculate $\sub{\super{D}{II}}{KL}$ using Equation \eqref{eq:dkl2} for each $\sub{\vect{a}}{trial}$ compared to $\vect{a}_0$. Finally, we use Equation \eqref{eq:kld2interp} to set confidence intervals by choosing appropriate contours of $\sub{\super{D}{II}}{KL}$. For example, if we had a Gaussian probability distribution, then the ``1-$\sigma$'' error bar corresponds to the level where the probability is $e^{-1/2} \approx 24$\% of its peak value. The equivalent in terms of the KLD is
\begin{equation}
\Big\langle \log \frac{ \mathcal{P}( \mathcal{H}_{\vect{a}_0} | \vect{J} ) }{ \mathcal{P}(\mathcal{H}_{\sub{\vect{a}}{trial}}|\vect{J}) }\Big\rangle_{\vect{J}} = \sub{\super{D}{II}}{KL} = 1/2.
\end{equation}
For the Gaussian, this level corresponds to 68 percent of the total probability, which is why it is customary to quote 68 percent confidence contours for non-Gaussian probability distributions. Likewise, the ``2-$\sigma$'' level, where parameters are $e^{-2}\approx 1/20$ as likely as the best fit, is $\sub{\super{D}{II}}{KL} = 2$, and corresponds to the  95-percent  confidence interval; $\sub{\super{D}{II}}{KL} = 4.5$ is equivalent to the 99-percent (``3-$\sigma$'') contour.

\section{Creating the mock stellar halo}
\label{sec:process}

We construct a mock stellar halo in a potential with analytic actions to test the method. For these initial trials, we use a mock halo that is entirely accreted; i.e., all the stars in the mock halo are tidal remnants of disrupted satellites. In this section we describe the potential (\ref{subsec:pot}), the mock-halo generation process (\ref{subsec:ics}), and our method for ``observing'' the halo with Gaia by convolving the observables with Gaia-like errors (\ref{subsec:conv}).

\subsection{Potential and input parameters}
\label{subsec:pot}
We choose the isochrone potential
\begin{equation}
 \Phi(r) = -\frac{M}{b+\sqrt{r^2+b^2}}
\end{equation} 
 because it has analytic expresssions for the actions $(J_r,L,L_z)$, where $L$ is the absolute value of the total angular momentum, $L_z$ is its $z$ component, and $J_r$ is the radial action
\begin{equation}
 J_r = \frac{G M}{\sqrt{-2E}} - \half \left(L+\sqrt{L^2 + 4 G M b}\right).
\end{equation}
The specific energy $E$ is given by the standard expression
\begin{equation}
 E = \half \vect{v}\cdot\vect{v} + \Phi(r).
\end{equation} 
The isochrone potential is actually not a very good match to what we know about the shape of the real Milky Way or about simulated cosmological dark matter halos. In particular, this potential is spherically symmetric, whereas the Milky Way has a disk and its halo may be flattened as well \citep{2010ApJ...714..229L,2013ApJ...773L...4V}. In principle any potential for which the actions can be calculated will work; for simplicity we start with an analytically tractable example. We choose as input parameters $\sub{\vect{a}}{true}$ the scale length $b=8$ kpc and the total mass $M=2.7\times10^{12}\ M_{\odot}$ to roughly reproduce the mass \citep[e.g.][]{Sofue2012,Boylan-Kolchin2013,Bhattacharjee2014} and scale \citep{Battaglia2006,Deason2012,Kafle2012} of the Milky Way. However, with these parameters the circular velocity of the halo peaks at 17.6 kpc, where it is 500 km \unit{s}{-1}, declining to 315 km \unit{s}{-1} at 100 kpc (the edge of our mock halo); at the solar circle $v_c=420$ km \unit{s}{-1}. These velocities are significantly higher than the range measured for the Milky Way \citep{Deason2012,Kafle2012,Sofue2012,Bhattacharjee2014}, reflecting the fact that the isochrone profile is not a very good match; as a result of the shorter dynamical times, the streams in our mock halo will be more phase-mixed than streams of similar age in a more realistic Milky Way potential. Given that we consider only the streams' action-space distributions and discard phase information, we do not expect this difference to greatly affect our results.

\subsection{Making the streams}
\label{subsec:ics}
We construct a distribution of progenitor galaxies to mimic the basic features of the known Milky Way satellites, and integrate them in the galactic isochrone potential to make the accreted mock halo. For our purposes a progenitor ``satellite galaxy" is made up of equal-magnitude red-giant (RGB) stars following a spherical Plummer distribution. 

\begin{figure}
\begin{center}
\includegraphics[width=0.48\textwidth]{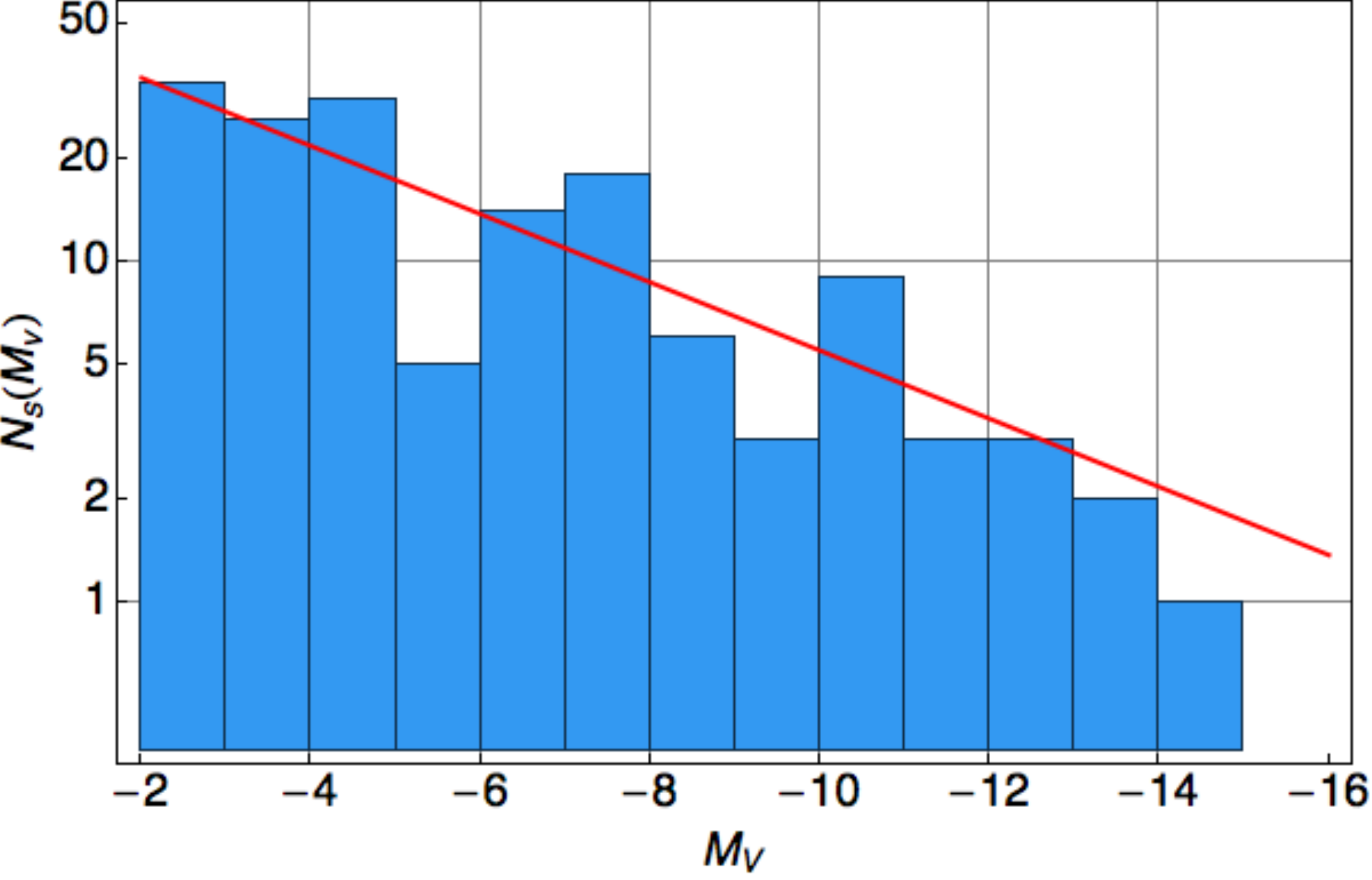}
\caption{The luminosities of the progenitors making up the mock halo (histogram; blue online) are drawn from the Koposov estimated luminosity function for known Milky Way satellite galaxes (solid line; red online).}
\label{fig:lumFunction}
\end{center}
\end{figure}

We begin by drawing the luminosities of the progenitors from the luminosity function of Milky Way dwarfs derived by \citet[][see Figure \ref{fig:lumFunction}]{Koposov2008}. From these luminosities we use the fundamental plane parameterized by \citet{Tollerud2011} to obtain the progenitors' half-light radii $\sub{r}{1/2}$ and total dynamical mass at this radius $\sub{M}{1/2}$. The half-light radius is related to the scale radius $r_s$ of the Plummer spheres representing the progenitors by 
\begin{equation}
 r_s = \sub{r}{1/2}\sqrt{2^{2/3}-1}.
\end{equation}
To determine the number of stars $n_*$ in a progenitor we assume a mass-to-light ratio of 2 for the \emph{stellar} component and count 1 RGB star per 40 solar masses of stars \citep{Marigo2008,Helmi2011}. Thus $ n_* = (2L)/40 = L/20$. We require a minimum of 20 red giants per galaxy, corresponding to a minimum luminosity of $4 \times 10^2\ L_{\odot}$. We sample the Plummer DF with the given $r_s$ and total mass $n_*$ times to get the positions and velocities of the RGB stars in each progenitor relative to its center of mass.

Next we select an orbit for each progenitor. We first determine its orbital energy by choosing an apocenter distance, $r_a$. We would like the average density profile of the stars in the mock halo to be consistent with observations of the stellar halo, which has $\rho \propto r^{-3.5}$ \citep{Helmi2002}, so we select the apocenters of the center-of-mass orbits from a probability distribution in $r_a$ that corresponds to this density distribution. Given that
\begin{equation}
 p(r_a) dV = \rho(r_a) dV \propto r_a^{-3.5} dV,
\end{equation}
and using $dV=4\pi r_a^2 dr_a$, we find that the normalized probability distribution for $r_a$ is
\begin{equation}
 p(r_a) dr_a = \half \sqrt{\frac{r_c}{r_a^3}} dr_a, 
\end{equation} 
where we have calculated the normalization by setting an inner cutoff radius $r_c=0.3$ kpc (inside the bulge) to avoid the divergence of the distribution as $r_a\to0$. We draw the apocenters from this probability distribution, throwing out values of the apocenter distance outside the range $3<r_a<100$ kpc to focus on streams at locations accessible to observations. To mimic the mass-dependent effect of dynamical friction, we assign smaller $r_a$ values to progenitors with larger $M$ with some scatter, and explicitly limit the apocenter distances of the six largest progenitors to within 10 kpc. The initial angular position of each progenitor $(\theta, \phi)$ is chosen from a uniform distribution on $(-1<\cos \theta<1, 0<\phi<2\pi)$.

\begin{figure*}
\begin{center}
\includegraphics[width=\textwidth]{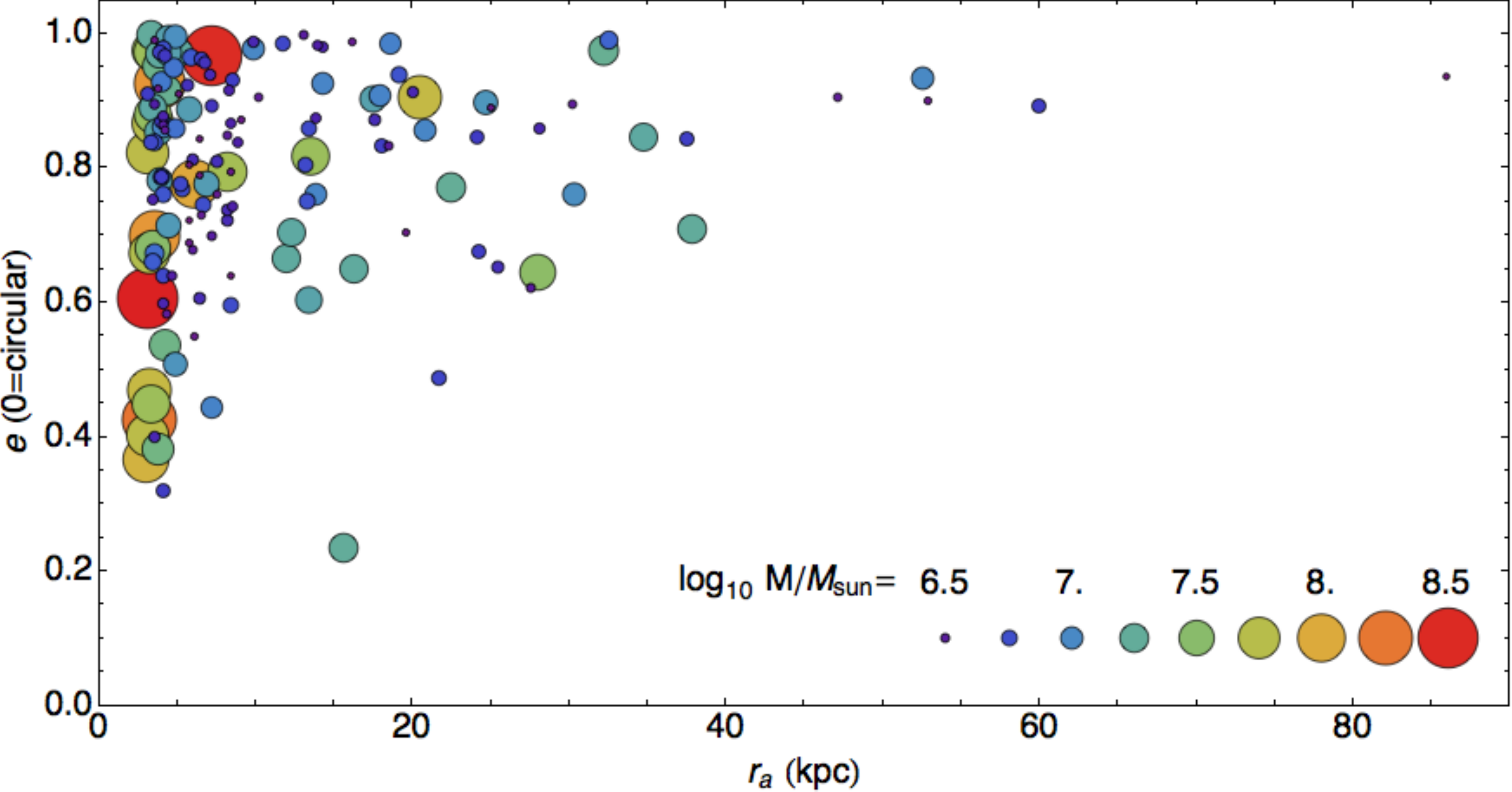}
\caption{Masses, eccentricities, and apocenter radii of the progenitors making up the mock halo.}
\label{fig:orbits}
\end{center}
\end{figure*}

We determine the orbital angular momentum by choosing a circularity $\eta \equiv L/L_c(r_a)$ based on the distribution found by \citet{Wetzel2011} for infalling progenitors at the virial radius of Milky-Way-mass halos at $z=0$. The orbit inclination $\cos i = L_z/L$ determines the fraction of the angular momentum in the $z$ direction and the direction of the orbit; we choose $|\cos i|$ uniform on $0<|\cos i|<\sin \theta$, and then choose the sign of $\cos i$ to be positive or negative with equal probability. The orbital properties of the progenitors are shown in Figure \ref{fig:orbits}.

The last step is to choose an infall time for each progenitor. We choose a random number of radial orbital periods from a continuous uniform distribution between 5 and 350, with the stipulation that the total orbital time cannot be longer than 13.6 Gyr (350 radial periods equals a Hubble time for the progenitor with the smallest apocenter). The sampled stars are integrated as test particles in the galactic potential for the chosen time, and the process is repeated for each progenitor to build up the mock halo.

\begin{figure}
\begin{center}
\includegraphics[width=0.48\textwidth]{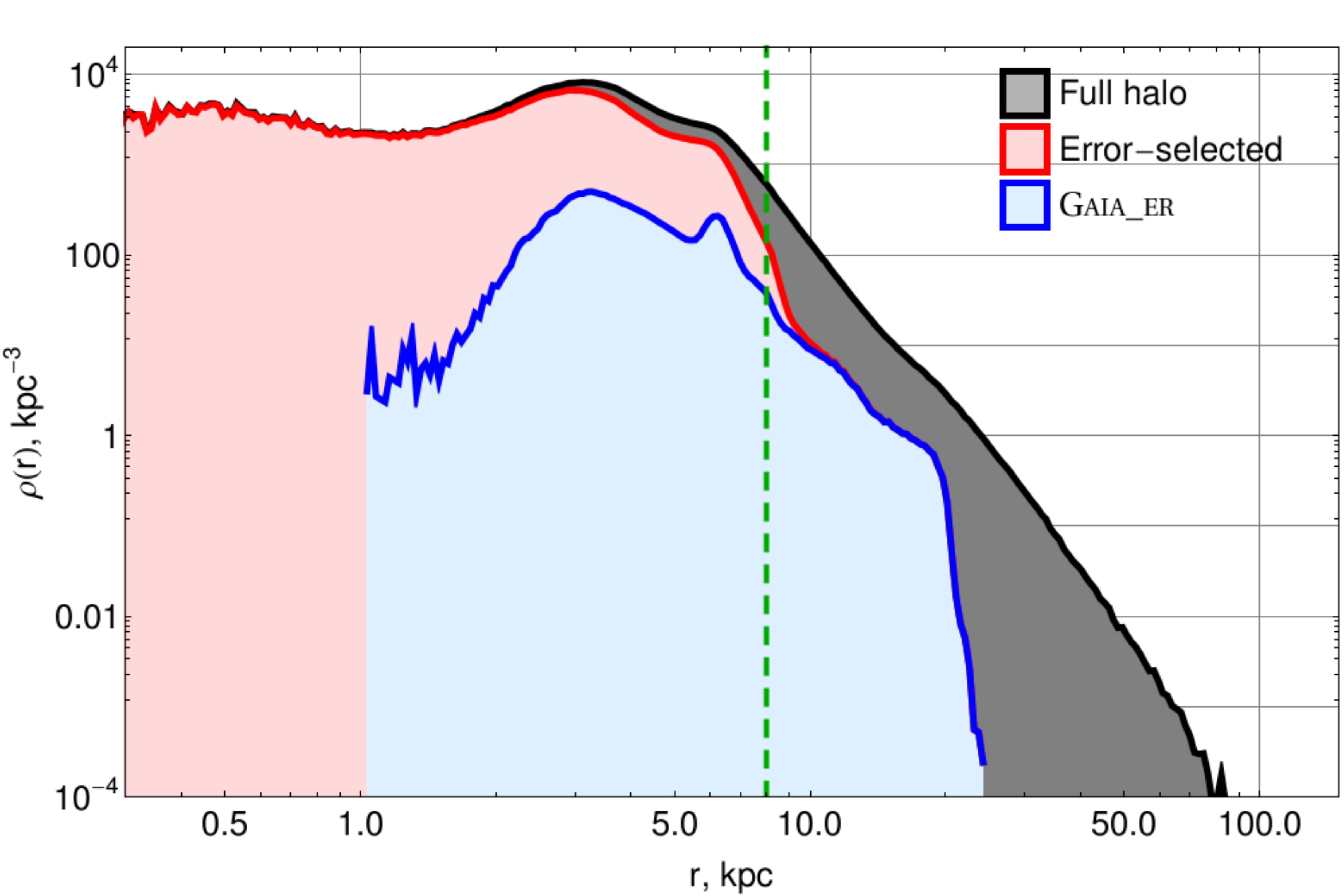}
\caption{Galactocentric radial density distribution of stars in the mock halo, constructed from the error-convolved parallaxes. The background curve (black online) shows the distribution for all stars; the first overlapping curve (red online) shows the distribution of the error-selected stars (see Section \ref{subsec:conv}); the foremost curve (blue online) shows the distribution of stars after selection for both error and ``energy'' ({\sc Gaia\_er}; see Section \ref{subsec:streamsel}). The green dashed vertical line at 8 kpc marks the scale radius of the input potential.}
\label{fig:rdist}
\end{center}
\end{figure}

\begin{figure}
\begin{center}
\includegraphics[width=0.48\textwidth]{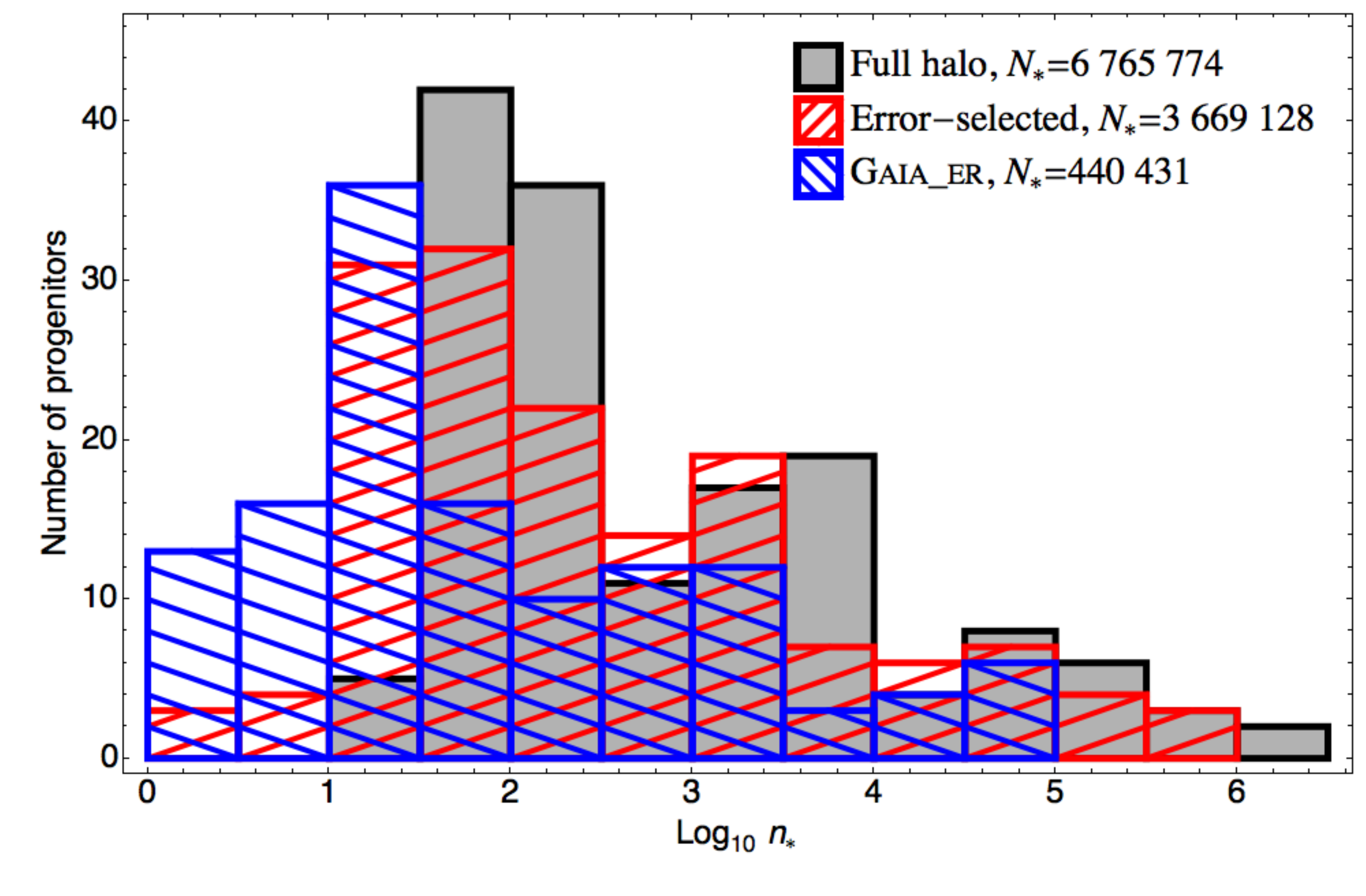}
\caption{Distribution of the number of stars $n_*$ in each progenitor, for the entire mock halo (black filled with gray), the error-selected sample (hashed; red online), and the energy- and error-selected sample ({\sc Gaia\_er}, reverse-hashed; blue online). The error-selected sample includes stars from all but one of the 153 progenitors in the full mock halo; {\sc Gaia\_er} contains stars from 128 progenitors.}
\label{fig:nstarstream}
\end{center}
\end{figure}

Estimates based on cosmological simulations and semi-analytic modeling project that the Milky Way should contain about 110 thin streams (disrupted progenitors with stellar mass less than $10^5\ M_\odot$) with more than 20 RGB stars brighter than 20th magnitude \citep[][p. 14]{chiappini2013}. In this work we assume an absolute magnitude $M_V=1$ for all the stars in the stream, so this magnitude cut translates to a cut in heliocentric distance of about $d_\odot \lesssim 60$ kpc, or a cut in galactocentric distance of just under 100 kpc (Figure \ref{fig:rdist}, gray curve). Choosing 153 progenitors and integrating their orbits as described above results in a Milky Way stellar halo with the right number of thin streams in this distance range. A summary of the bulk properties of the mock halo is given on the first line of Table \ref{tbl:sampleStats}, while the distribution of progenitor sizes is shown in gray in Figure \ref{fig:nstarstream}. Although we generate progenitors with orbit apocenters as large as 100 kpc, the small average galactocentric distance of the halo (about 6 kpc) indicates that the stars are very centrally concentrated. This is both because the radial distribution we use is quite steep, and because we preferentially placed the 5 or so largest satellites on fairly small orbits.

\begin{figure*}
\begin{center}
\plottwo{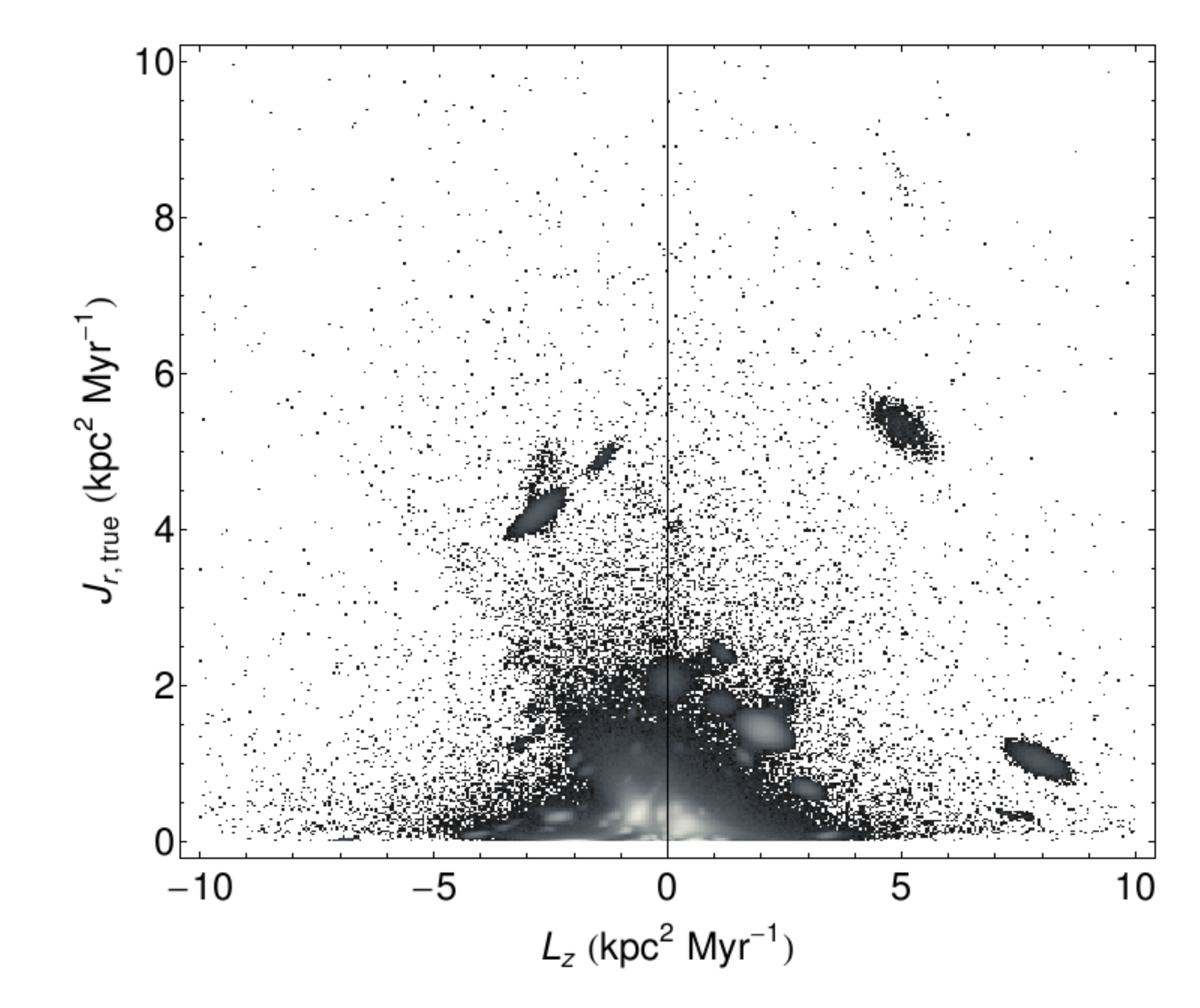}{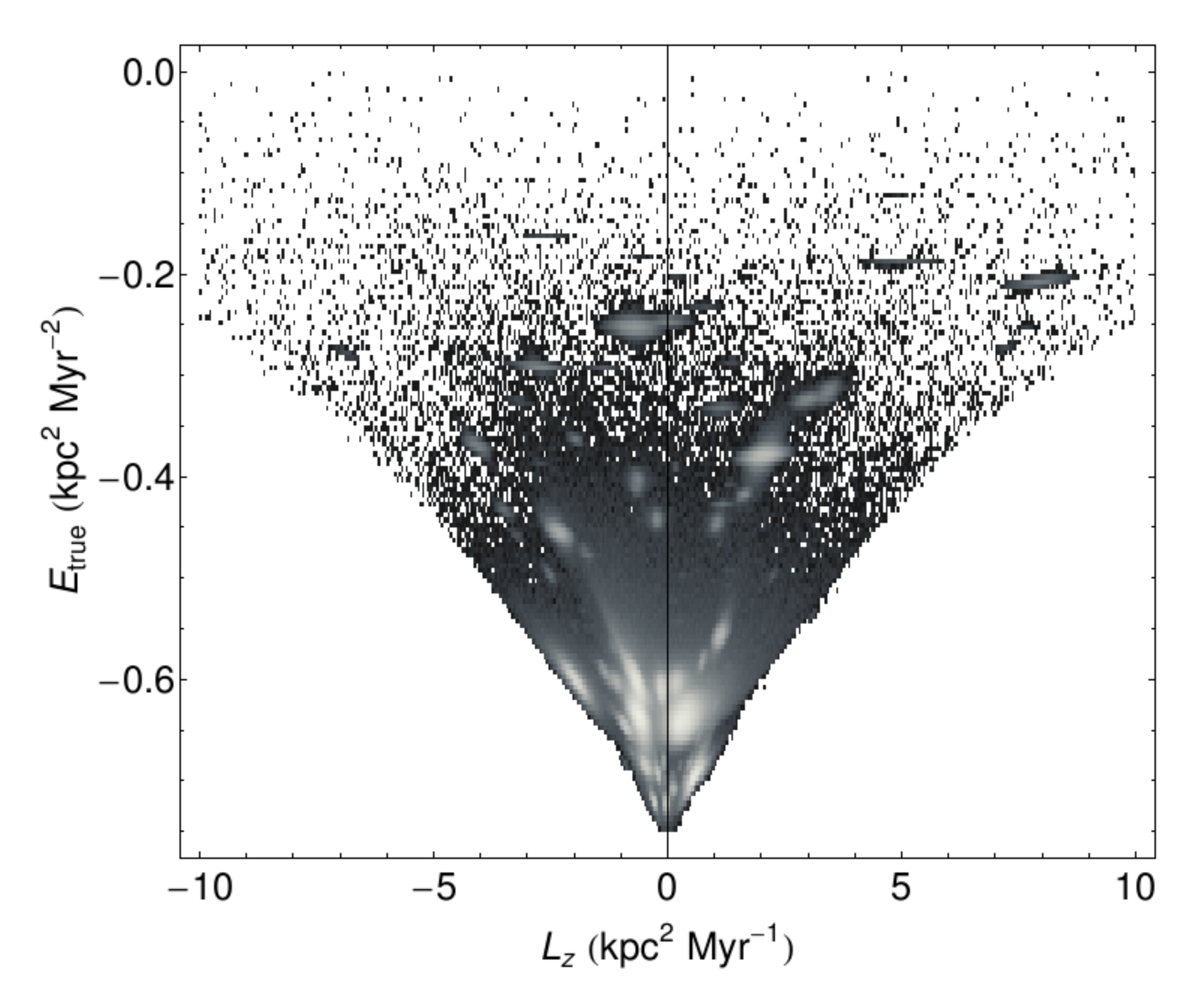}
\caption{Log density in $J_r-L_z$ space (left) and $E-L_z$ space (right) of all stars in the mock halo (lighter gray = more stars). $J_r$ and $E$ were calculated using the input potential.}
\label{fig:trueActions}
\end{center}
\end{figure*}

Figure \ref{fig:trueActions} shows that the action-space distribution of the stars in the mock halo is indeed very clumpy. At larger values of the actions the clumps are well-separated and distinct, while at smaller actions, which roughly correspond to orbits deeper in the potential, the clumps begin to overlap and blend together. These inner structures are also more likely to be well-mixed in $x$ and $v$, since the radial and azimuthal frequencies are, roughly speaking, inversely proportional to the actions.

\begin{table}
\caption{Mock halo statistics}
 \begin{tabular}{llllll}
& & $\langle d \rangle$, & $\langle d_\odot \rangle$, & &  $\sub{N}{pro}$, \\
Sample & $N_*$ &  kpc &  kpc & $\sub{N}{pro}$ & $n_*>100$ \\
\hline
\hline
Full mock halo & 6 765 774 & 6.20 & 10.4 & 153 & 106 \\
After error selection & 3 669 128 & 4.84 & 8.26 & 152 & \phantom{0}82 \\
{\sc Gaia\_er} & \phantom{0 }440 431 & 6.51 & 7.96 & 128 & \phantom{0}47 \\
\end{tabular}
\tablecomments{This table summarizes the contents of the full mock halo (Section \ref{subsec:ics}), the part selected for data completeness and quality after error convolution (Section \ref{subsec:conv}), and the energy-selected sample used for the fit (Section \ref{subsec:streamsel}). Listed are the total number of stars in each sample $N_*$, mean galactocentric distance $\langle d \rangle$, mean heliocentric distance$\langle d_\odot \rangle$, number of progenitors \sub{N}{pro}, and number of progenitors with more than 100 stars. }
\label{tbl:sampleStats}
\end{table}

\subsection{Observables and noise convolution}
\label{subsec:conv}
We add noise to the data based on the projected performance of the Gaia survey \citep{DeBruijne2012}. To simulate observational errors we calculate the parallax, proper motions, line-of-sight velocity (RV), and sky positions and determine the expected error on each observable based on the current error models for Gaia\footnote{see \url{http://www.rssd.esa.int/index.php?page=Science_Performance\&project=GAIA}}. We set an upper limit of 20\% for the expected relative parallax error, consistent with estimates for the photometric parallax error for RGB stars. To calculate the errors it is necessary to assign an absolute magnitude and spectral class to each star in the stream; for this work we assume all the stars are red KIII giants with $M_V = 1$. Each observable for each star is then drawn from a one-dimensional Gaussian distribution centered on the true value with the width of the expected error, assuming that the errors on the observables are uncorrelated.  The convolved observables are then converted back into ``noisy'' 6D positions $(\tilde{\vect{x}}_i,\tilde{\vect{v}}_i)$. We then select all the stars with full six-dimensional phase space coordinates: effectively, this means those that have radial velocity (RV) measurements. For KIII metal-poor stars, this corresponds to a maximum $V$ magnitude of 17.29.  We also throw out stars whose predicted transverse velocity error is larger than the maximum predicted RV error of 18.15 km \unit{s}{-1}, so that errors on all components of the velocity are comparably sized. These cuts in data quality eliminate 46 percent of the stars from the sample, as shown in the second line of Table \ref{tbl:sampleStats}, and remove mostly stars at galactocentric distances larger than 25 kpc (Figure \ref{fig:rdist}, red curve). The effect is seen in the average galactocentric and heliocentric distances, which are both significantly smaller than for the full halo.

\section{Stream selection}
\label{subsec:streamsel}
One strength of our fitting method is that it does not require a priori knowledge of which stars belong to which stream; the KL divergence simply measures the total clustering of all the stars in the sample. The only requirement is that the different action-space clumps are sufficiently separated in action space that the statistic can distinguish the clumpiness of the distribution. Because the KL divergence measures the total clustering, it works best when the clump size is small compared to inter-clump separation, especially in dimensions that depend on the potential. Deeper in the potential the various streams can overlap each other enough that the distribution is not easily distinguishable from a smooth one; this region must be removed from the sample. To remove it, we make an informed guess for the potential of the Milky Way: an isothermal sphere (logarithmic potential) with constant circular velocity $v_c$ and some offset $\phi_0$ chosen so that all the stars in the sample are bound:
\begin{equation}
\label{eq:trialPotential}
\phi(r) = v_c^2 \ln r - \phi_0.
\end{equation}
For the circular velocity $v_c$, we use the average value of $|\vec{L}|/r$ for all the error-selected stars in the error-convolved sample, which is 249 km \unit{s}{-1}. We set the offset $\phi_0$ conservatively to
\begin{equation}
 \phi_0 \equiv \textrm{max}\left[ v_c^2 \ln r \right] + \textrm{max}\left[\half \vec{v}\cdot\vec{v}\right],
\end{equation}
which works out to $0.78$ \unit{kpc}{2} \unit{Myr}{-2}, to guarantee that all the stars are bound.

\begin{figure}
\includegraphics[width=0.45\textwidth]{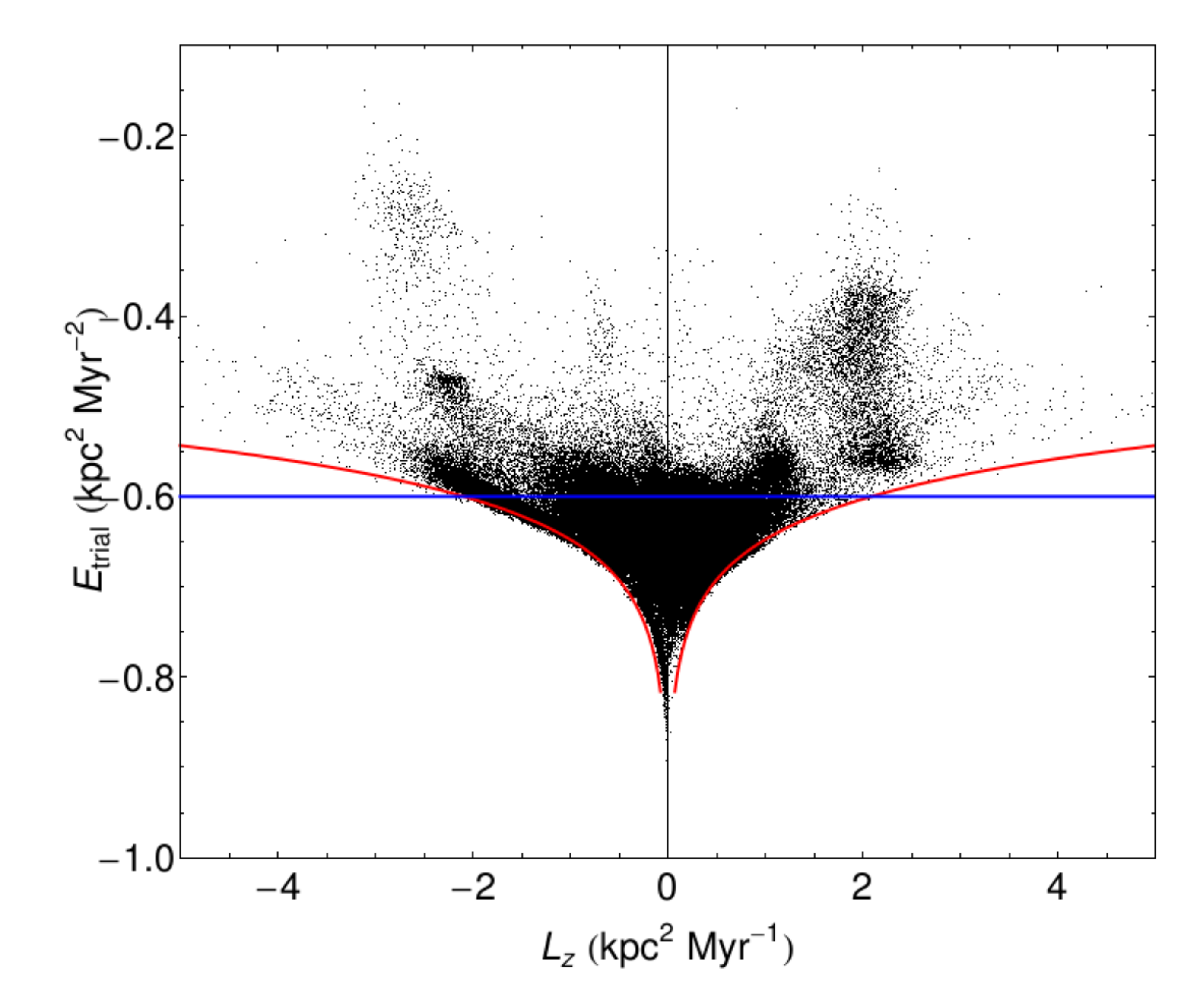}
\caption{Stars from the mock halo are selected for the fit by defining an energy cut based on visible clustering in the space $(L_z,\sub{E}{trial})$. Energy is calculated assuming the trial potential of Equation \eqref{eq:trialPotential}. We used an energy cutoff $\super{\sub{E}{trial}}{min}=-0.6\ \unit{kpc}{2}\unit{Myr}{-2}$ (horizontal line; shown in blue online) low enough to include a few streams with quasi-circular orbits (those near the bounding curve; shown in red online). Stars above the horizontal (blue) line are included in the fit. One-tenth of the stars are plotted here. \label{fig:select}}
\end{figure}

Using this guess for the potential, we calculate an energy $\sub{E}{trial}$ for all the stars, using their error-convolved 6D positions, and plot it against the calculated $z$ component of the angular momentum, $L_z$, as shown in Figure \ref{fig:select}. Even though this is not the correct energy there is still a fair amount of clustering visible in this space, especially at values of $\sub{E}{trial}$ significantly above the minimum for a given $L_z$ (solid red line). We assess the clustering in this plot by eye to choose a cutoff value $\super{\sub{E}{trial}}{min}$, selecting all the stars with $\sub{E}{trial}>\super{\sub{E}{trial}}{min}$. The goal is to include as many stars as possible while avoiding the region where the structures completely overlap. In order to include all of the well-separated clumps as well as some streams on more circular orbits (ones near the red line), we chose $\super{\sub{E}{trial}}{min}=-0.6\ \unit{kpc}{2}\unit{Myr}{-2}$ (blue line) after some experimentation. We refer to this sample as {\sc Gaia\_er}. 

Applying this approximate energy cutoff selects about $5\times10^5$ stars, about 12\% of the error-selected sample and about 6.5\% of the stars in our whole mock stellar halo (see the third line of Table \ref{tbl:sampleStats}). Figure \ref{fig:rdist} (blue curve) shows that this selection indeed cuts out mostly stars at small galactocentric distances, deep within the halo. This is also apparent from the increase in the average galactocentric distance, which is now nearly equal to the average heliocentric distance of the stars in this sample. Figure \ref{fig:nstarstream} shows how the error and energy selections affect the number of stars per progenitor and the total number of progenitors represented in the halo. 

\begin{figure*}
\begin{center}
\plottwo{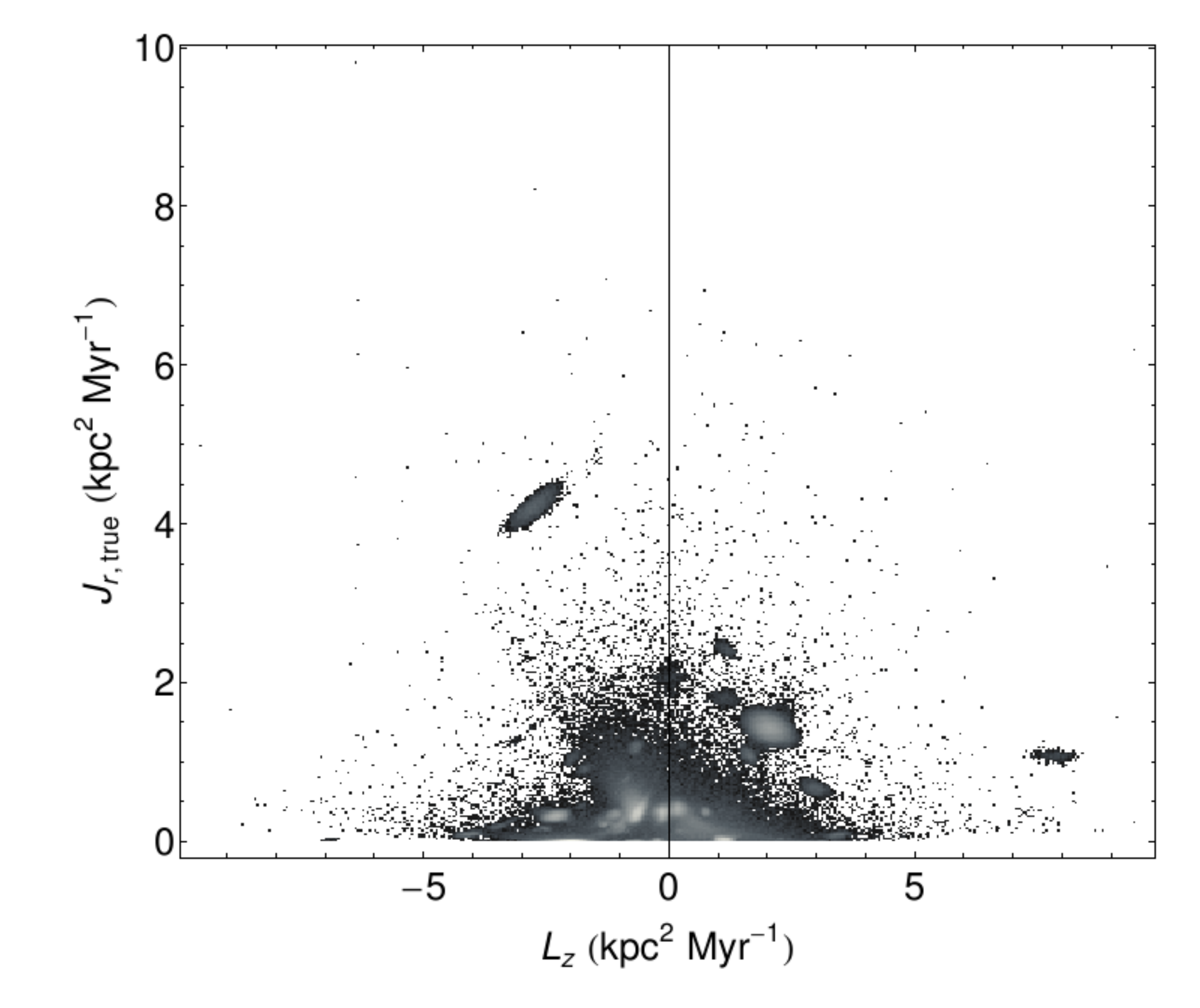}{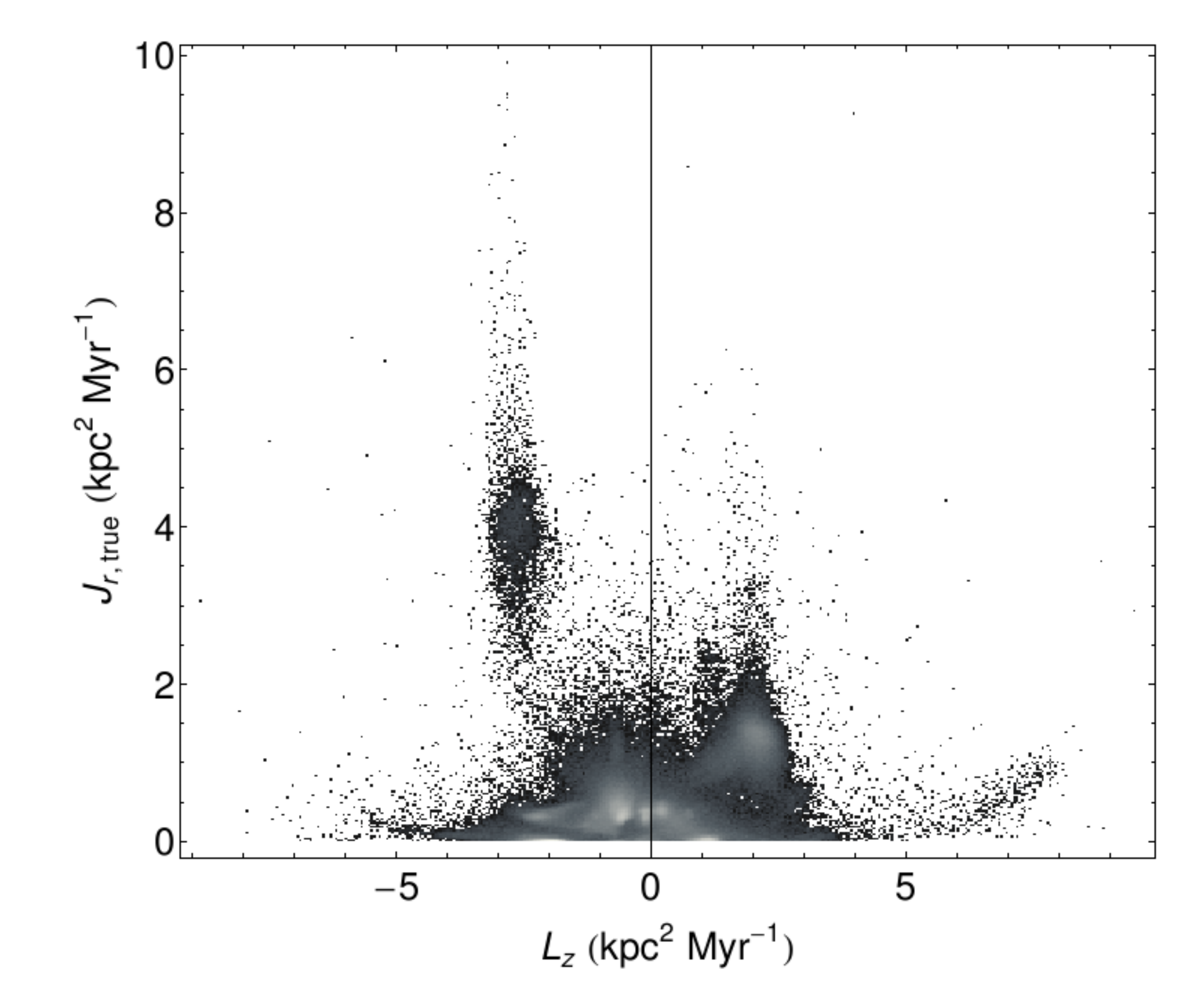}
\caption{Distribution (log density) of stars in action space (as in the left panel of Figure \ref{fig:trueActions}) for the samples {\sc Gaia\_ne} (without errors; left) and {\sc Gaia\_er} (with errors; right).}
\label{fig:ActionsAfterCuts}
\end{center}
\end{figure*}

We also recorded the ``true'' positions and velocities of the stars selected with this energy criterion, prior to error convolution, to isolate the effect of the observational errors. We call this sample {\sc Gaia\_ne}. Viewed in action space (Figure \ref{fig:ActionsAfterCuts}), we see that although the errors do blur the existing substructure, they do not destroy all the information: many clumps are still present in {\sc Gaia\_er}. A comparison with the left panel of Figure \ref{fig:trueActions} shows that the effective distance cut imposed by the error selection has eliminated a few objects at large $J_r$ and $L_z$, while the energy cut has removed a fraction of the clumps near $J_r=L_z=0$ to thin out this region, as intended. 

\section{Computation of the KL divergence}
\label{subsec:klcomp}
The KLD we employ can be computed even if the function $p(\vect{x})$ is not known, provided that we have a sampling of $N_*$ points $\vect{x}_i$ (observed points) assumed to be drawn from $p(\vect{x})$.
In that case an ``observed'' distribution $\tilde{p}$ can be constructed (e.g., using a density estimator) from the observed $\vect{x}_i$ and used in place of the true distribution $p$. An observed $\tilde{q}$ can also be constructed from a set of points $\vect{y}_i$, again using a density estimator, as long as it is then evaluated at the points $\vect{x}_i$ drawn from $p$, which will not in general be equal to $\vect{y}_i$, to calculate the KLD. In practice, $\vect{x}_i$ and $\vect{y}_i$ correspond to the actions computed using different values of $\vect{a}$ (in the case of $D^{II}_{KL}$) or to the actions and shuffled actions with a given $\vect{a}$ (in the case of $D^{I}_{KL}$). This leads to explicit expressions for the Monte Carlo integrations of the two KLDs:
\begin{equation}
  D^{I}_{KL} = \frac{1}{N_*}\sum_{i=1}^{N_*} \log \frac{\tilde{f}_{\vect{a}}\left(\vect{J}_i(\vect{a})\right)}{\super{\tilde{f}_{\vect{a}}}{shuf}\left(\vect{J}_i(\vect{a})\right)}
\end{equation}
and
\begin{equation}
  D^{II}_{KL} = \frac{1}{N_*}\sum_{i=1}^{N_*} \log \frac{\tilde{f}_{\vect{a}_0}\left(\vect{J}_i(\vect{a}_0)\right)}{\tilde{f}_{\sub{\vect{a}}{trial}}\left(\vect{J}_i(\vect{a}_0)\right)},
\end{equation}
 where the $\tilde{f}$ are constructed using a density estimator from the actions $\vect{J}_i(\vect{a})$ of the stars in the data set, for various values of the parameters $\vect{a}$. The arbitrary $\vect{J}$ from the integrals in Equations \eqref{eq:dkl1} and \eqref{eq:dkl2} have now been replaced by the specific $\vect{J}_i(\vect{a})$ of the stars in the sample, so that the integral is sampled at points drawn from the correct distribution for Monte Carlo integration. 

Alternatively, we can calculate the KLD by explicitly discretizing the integral in each step, summing over some number of volume elements $\sub{N}{grid}$ with (possibly varying) sizes $\Delta_i$, and evaluating the probability density at the centers $\vect{J}_k$ of these volume elements: 
\begin{eqnarray}
  D^{I}_{KL} &=& \sum_{k=1}^{\sub{N}{grid}} \Delta_i\ \tilde{f}_{\vect{a}}(\vect{J}_k) \log \frac{\tilde{f}_{\vect{a}}(\vect{J}_k) }{\super{\tilde{f}_{\vect{a}}}{shuf}(\vect{J}_k)} 
\end{eqnarray}
and
\begin{eqnarray}
  D^{II}_{KL} &=& \sum_{k=1}^{\sub{N}{grid}} \Delta_i\ \tilde{f}_{\vect{a_0}}(\vect{J}_k)  \log \frac{\tilde{f}_{\vect{a_0}}(\vect{J}_k)}{\tilde{f}_{\sub{\vect{a}}{trial}}(\vect{J}_k)}.
\end{eqnarray}
Tests where we compared single bivariate Gaussians of different sizes found that for resolutions comparable to the number of particles in the individual clumps of our mock stellar halo (5000 points or more) this method produced a more accurate estimate of the KLD than Monte Carlo. 

We used the modified Breiman density estimator developed and tested by \citet{2011A&A...531A.114F} to construct representations of the various $\tilde{f}$ on identical three-dimensional regular grids (i.e. constant $\Delta$), and then calculated the KLD by summing over the grid squares. The modified Breiman method uses an adaptive Epanechnikov kernel density estimator in which the optimal kernel size is determined through a pilot, non-adaptive density estimate with the same kernel shape. We used a $512^3$ grid in $(L_z, L, J_r)$, based on the results of our convergence testing.

\begin{figure*}
\plottwo{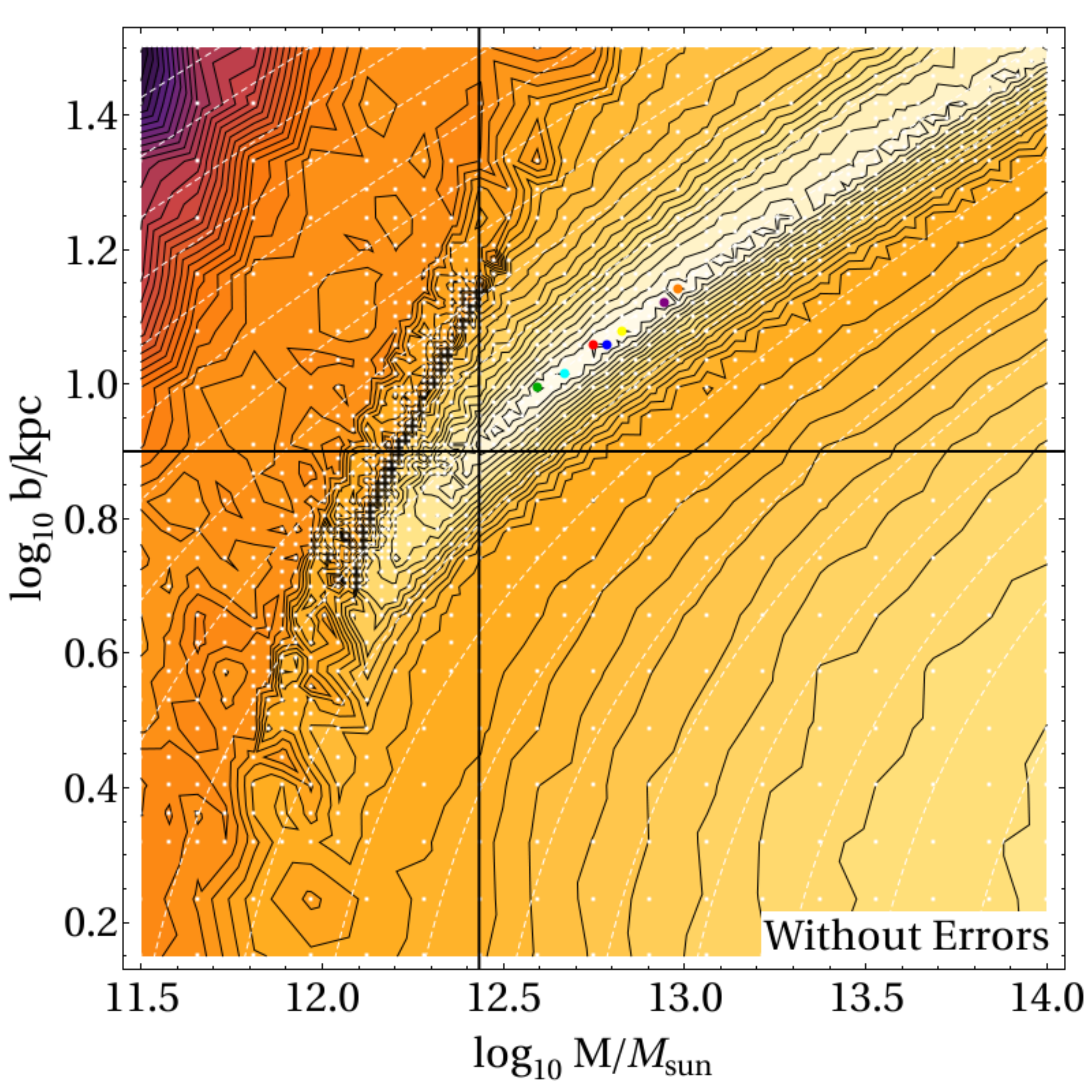}{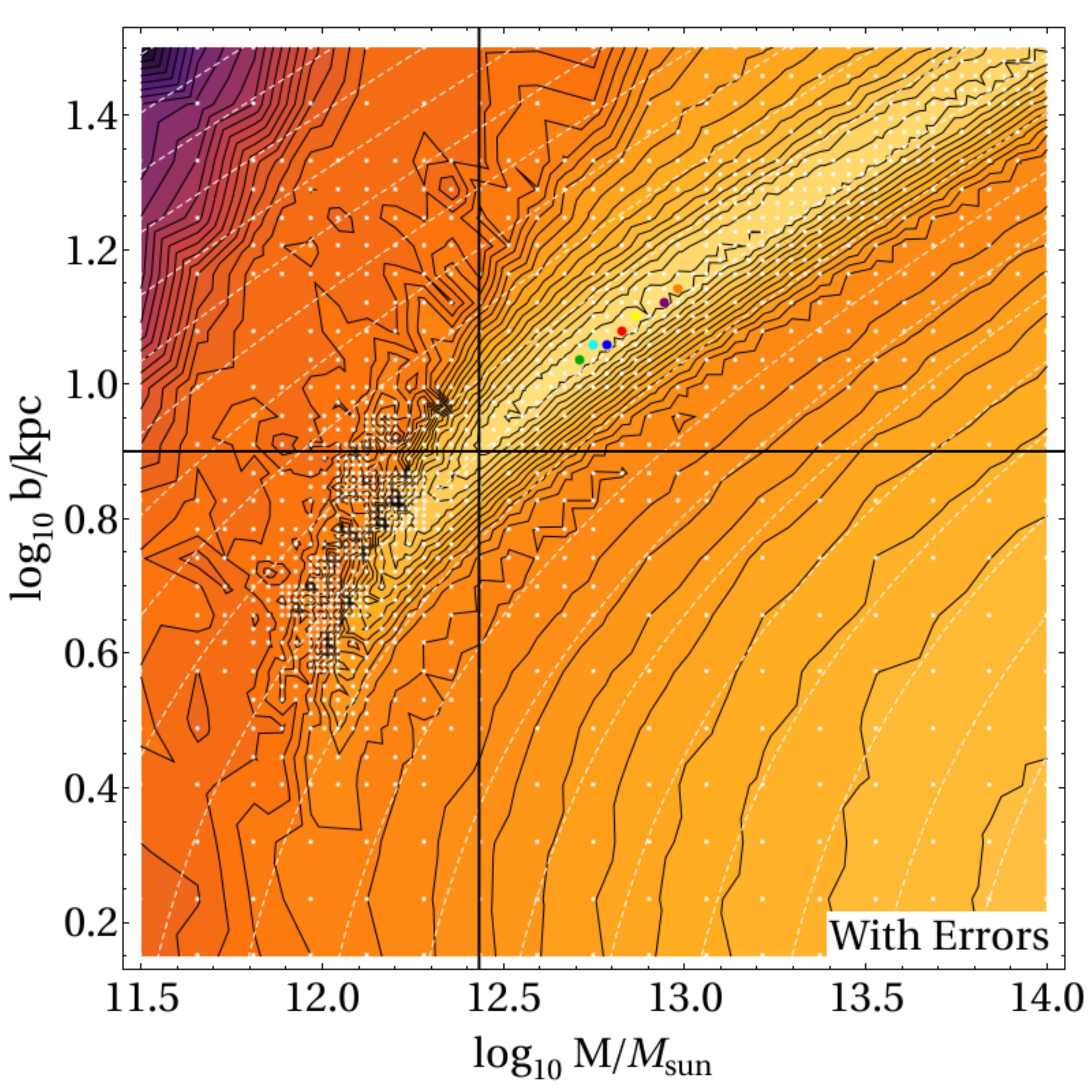}
\caption{Contours of $\sub{\super{D}{I}}{KL}$ for the samples {\sc Gaia\_ne} (left) and {\sc Gaia\_er} (right). The white points indicate sampled values of the parameters $\vect{a}\equiv (M,b)$. The black lines cross at $\sub{\vect{a}}{true}$; the red point is $\vect{a}_0$, the value at which $\sub{\super{D}{I}}{KL}$ is largest; the orange through purple points are the 2nd- through 7th-largest values of $\sub{\super{D}{I}}{KL}$. The contour spacing is 0.05 nats; the range is fixed to [0,2.45] nats in both plots (the color scale varies from dark purple at 0 to white at maximum). The white dashed lines are lines of constant $\sub{M}{enc}$ at the average galactocentric radius $\langle d_* \rangle$ (see Tables \ref{tbl:resultsEr} and \ref{tbl:resultsNe}). The value of $\sub{\super{D}{I}}{KL}$ at $\vect{a}_0$ is 2.41 in the left panel and 2.03 in the right panel.}
\label{fig:resultsStep1}
\end{figure*}

\begin{figure*}
\plottwo{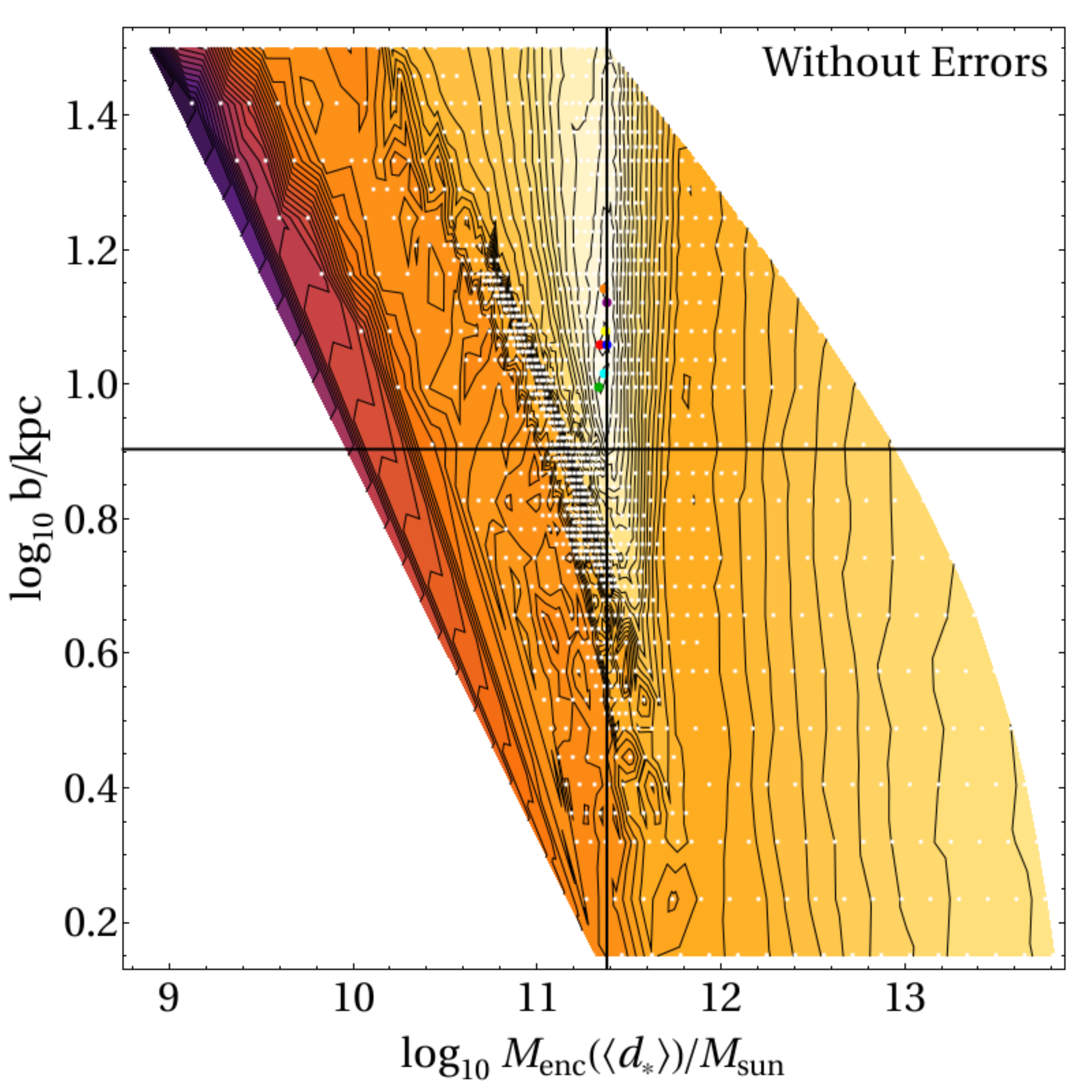}{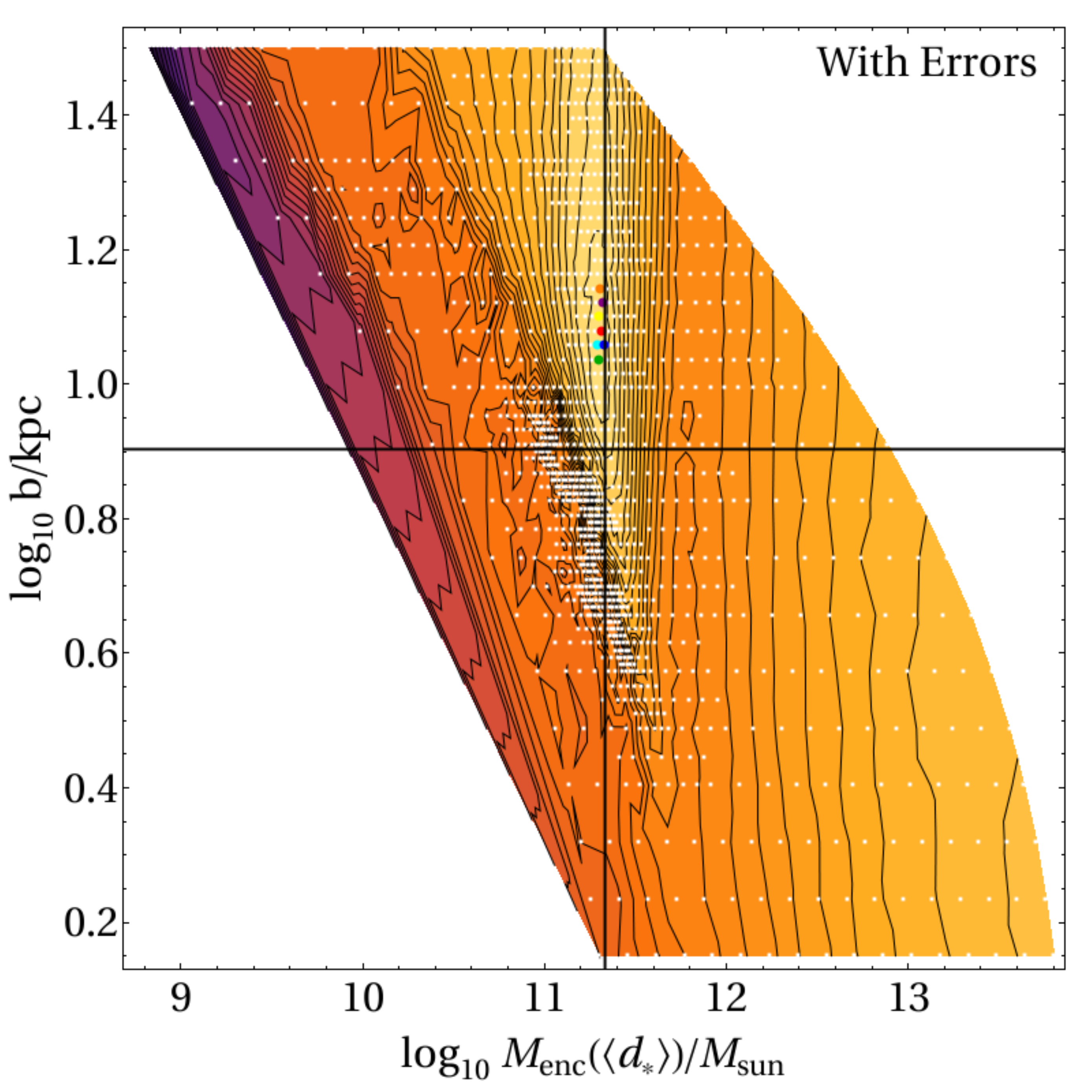}
\caption{As in Figure \ref{fig:resultsStep1}, but with the $M$-axis transformed to enclosed mass using Equation \eqref{eq:enclosedMass}.}
\label{fig:resultsStep1Menc}
\end{figure*}

For Step I, we find the best-fit parameters by calculating $\sub{\super{D}{I}}{KL}$ for different trial values of the potential parameters $\vect{a}$ on the intervals $[0.15,1.5]$ in $\log (b/\textrm{kpc})$ and $[11.5,14.0]$ in $\log (M/M_\odot)$. Starting with a 9x9 grid over this space, we use the {\sc HyperQuadTree} resampling code, kindly made available by Maarten Breddels, to progressively refine regions where the function is rapidly changing by adding new points at half the current spacing in each dimension. We continue refining until the 7 parameter combinations with the highest KLD values have mostly converged on the same peak; in practice this takes 4-5 iterations for a minimum spacing of about 0.01 dex in each parameter.

For Step II, we calculate $\sub{\super{D}{II}}{KL}$ for the full set of points sampled to maximize $\sub{\super{D}{I}}{KL}$, which are naturally focused on the region around the maximum. We keep track of the parameter-space volume associated with each sampling point. Then we use Equation \eqref{eq:kld2interp} to draw contours at $\sub{\super{D}{II}}{KL}=(1/2,2,9/2)$: as described in Section \ref{sec:kld} these contours are the relative probabilities at $(1\sigma, 2\sigma, 3\sigma)$ from the mean in a Gaussian distribution. In the next few sections, when discussing the results of Step II, we show plots of the normalized likelihood and superimpose these three confidence intervals.

We made two technical modifications to the calculation of the Step II KLD to eliminate sources of numerical noise. The first modification is to restrict the integration region of the KLD for this step to the range of actions in the distribution produced with the best-fit parameters, rather than the range including all stars in both the best-fit and trial distributions. This avoids the problem of large outliers in $J_r$, which is susceptible to numerical roundoff error since the density estimation grid has a fixed number of points in each dimension, and so for an extremely large range in $J_r$ will under-resolve the region containing most of the significant differences between distributions. Using a grid with fixed spacing instead is limited by memory constraints and for large outliers results in huge oversampled regions. Tests of this modification indicate that it does not significantly change the KLD values around the best fit, but does eliminate a region of spurious high probability (low KLD values).  

The second technical modification is to set a density ``floor" of $10^{-9}$ \unit{kpc}{-3} for grid cells that have a formal kernel estimate of zero density (that is, there are no stars within a few kernel widths of the cell). The purpose of this floor is to permit the calculation of the KLD for trial distributions that barely overlap with the best-fit distribution at all: cells that have nonzero density in the best-fit distribution but zero density in the trial distribution would otherwise have an undefined contribution to the KLD. If these are ignored, barely-overlapping distributions have vastly inflated probabilities.

We also had to deal with the different $J_r$ ranges of the two action distributions. As the mass and scale radius change, the total allowed range of $J_r$ values changes substantially, with an overall scaling proportional to $\sqrt{Mb}$. This stretching increases the size of all the action-space clumps in the distribution by the same factor, so if it is not compensated then potentials with larger $M$ and $b$ will be systematically disfavored since they produce lower-density clumps. Furthermore, we are interested in comparing the intrinsic clumpiness of the two distributions regardless of this relative scale, so we consider distributions in $\vect{J}=(L_z, L, J_r/\sqrt{Mb})$ rather than $\vect{J}=(L_z, L, J_r)$. This recenters the clumps close to one another in action space even for distributions with very different $M$ and $b$ so that the KLD calculation is primarily comparing the expansion and contraction of corresponding clumps, as desired. Performing this scaling also makes the volumes spanned by each distribution more similar, thus mitigating the effect of the limits we put on the integration volume for numerical reasons.

\section{Results from the full mock halo}
\label{sec:results}

In this section we discuss the results of the fit using the full mock halo samples {\sc Gaia\_er} and {\sc Gaia\_ne}, selected through the procedure described in Section \ref{subsec:streamsel}, with and without convolution with Gaia errors respectively.

\subsection{Finding the best-fit value (Step I)}
\label{subsec:step1results}

\begin{figure*}
\plottwo{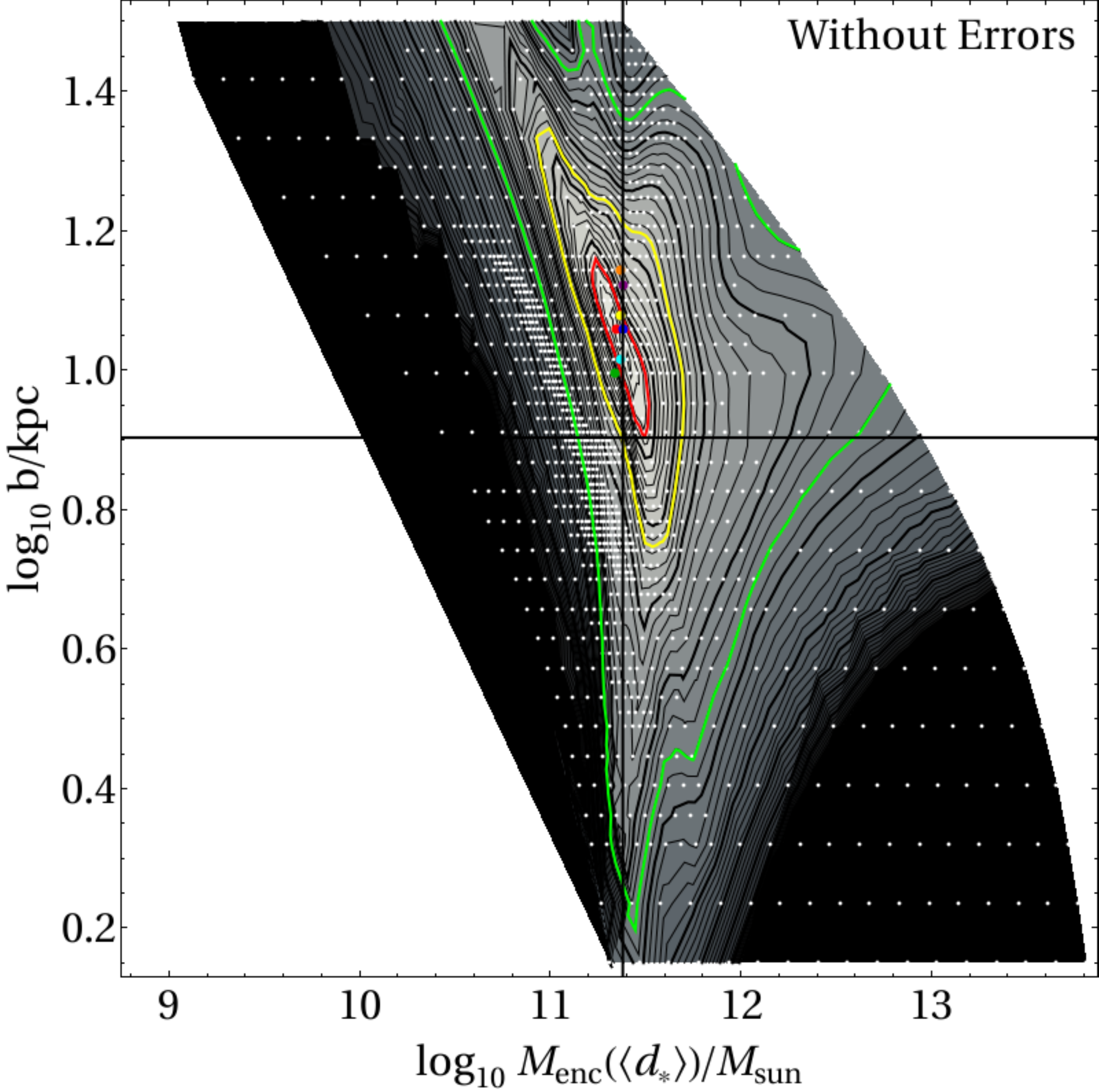}{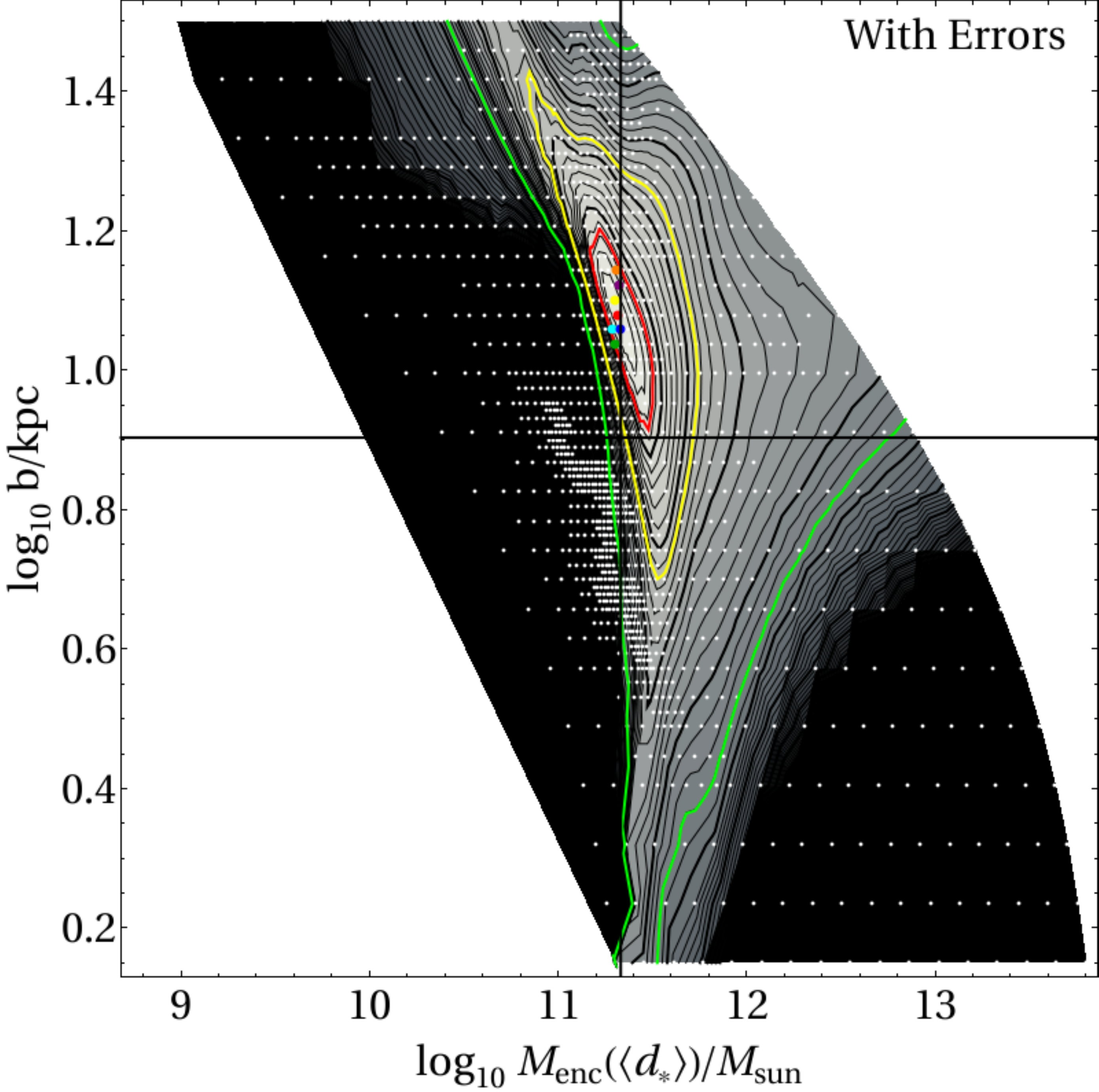}
\caption{Contours of $\sub{\super{D}{II}}{KL}$, the expected difference in log (base $e$) posterior probability with the best fit model (see Equations \ref{eq:dkl2} and \ref{eq:kld2interp}), for {\sc Gaia\_ne} (left) and {\sc Gaia\_er} (right). The white points are the sampled parameter values. Thin contours are 0.2 nats apart and thick contours are 1 nat apart, with range $0 \leq  \sub{\super{D}{II}}{KL} < 10$; regions with $\sub{\super{D}{II}}{KL} > 10$ have relative posterior probability less than $10^{-4}$ and are shown in black. Red, yellow, and green contours show $\sub{\super{D}{II}}{KL} = 1/2, 2, 9/2$ respectively (discussed in Section \ref{sec:kld}). The $M$-axis has been transformed to $\sub{M}{enc}$ using Equation \eqref{eq:enclosedMass}. }
\label{fig:resultsStep2}
\end{figure*}

Figure \ref{fig:resultsStep1} shows contours of the value of $\sub{\super{D}{I}}{KL}$ (Equation \ref{eq:dkl1}) as a function of the parameters for the samples {\sc Gaia\_er} (right) and {\sc Gaia\_ne} (left). The errors serve to decrease the contrast in the fit landscape by a bit less than half a ``nat'' (factor of $e$, since we use a natural logarithm to calculate the KLD) but do not appreciably change the bias between the best fit and the input value.  The best-fit parameters lie along a ridge of high $\sub{D}{KL}$ that corresponds to a measurement of the enclosed mass within the average radius of all the stars in the fitted sample, which for the isochrone case is
\begin{equation}
\label{eq:enclosedMass}
\sub{M}{enc}(\langle d_* \rangle) \equiv \frac{M \langle d_* \rangle^3}{\sqrt{b^2 + \langle d_* \rangle^2}\left(b^2 + \sqrt{b^2 + \langle d_* \rangle^2}\right)^2}.
\end{equation}
The dashed white lines in Figure \ref{fig:resultsStep1} are contours of constant $\sub{M}{enc}$ for $\langle d_* \rangle = 6.85$ (the average distance of the stars with no errors, left-hand panel) or $\langle d_* \rangle = 6.51$ (average distance after error convolution, right-hand panel). The orbits of the individual streams can be thought of as each contributing a measurement of the enclosed mass at a different average radius. These individual estimates then overlap in some region where the best-fit is found, so choosing slightly different sets of streams can move the overlap region along the ridge, with a more heterogeneous orbit selection giving a better result. 

If we calculate $\sub{M}{enc}(\langle d_* \rangle)$ for each sampled point in parameter space using Equation \ref{eq:enclosedMass} and redraw the KLD contours in the space $( \sub{M}{enc}(\langle d_* \rangle), b)$, then the ridge of high KLD becomes a narrow range in enclosed mass, as shown in Figure \ref{fig:resultsStep1Menc}. This indicates that the enclosed mass should be recovered very well, but that the scale radius will be more difficult to estimate. As we will discuss shortly, this is primarily because of the limited range in galactocentric distances imposed by the magnitude limit for Gaia's radial velocities---as shown in Figure \ref{fig:rdist}, this range is equivalent to only about three scale radii.

The secondary maximum in the lower right corner of the parameter space in both panels of Figure \ref{fig:resultsStep1} represents a deep point-mass-like potential. This solution is asymptotic in $\sub{D}{KL}$: if the action-space distribution is smooth, the overall density can grow by increasing the mass and decreasing the scale radius, collapsing all the orbits deeper into the trial potential. As more of action space is filled with merged or overlapping clumps, this solution starts to dominate, causing the KLD to increase without bound in this direction in parameter space, while the desired solution at nonzero $b$ becomes a secondary local maximum. The best strategy we found to avoid this asymptotic region is to gradually lower the energy threshold used for the selection and compare the best-fit results as more and more stars are added to the data set used for the fit.

\subsection{Setting error bounds on the best-fit value (Step II)}
\label{subsec:step2results}

 Figure \ref{fig:resultsStep2} shows the result of Step II, with (right) and without (left) Gaia errors. Recall that in Step II we use the KLD (Equation \ref{eq:dkl2}) to compare the distribution of actions at the best-fit values identified in Step I with values at the other points in the parameter grid. As discussed in Section \ref{sec:kld}, this version of the KLD is interpreted probabilistically, so the contours in these two figures show the expectation value of the difference in log posterior probability between the best-fit and other values of the parameters (Equation \ref{eq:kld2interp}). The confidence contour shown in red (yellow, green) is $\sub{\super{D}{II}}{KL}=1/2$ 
(2,$9/2$). As discussed in Section \ref{sec:kld} these levels correspond approximately to 68, 95, and 99\% confidence contours if the probability surface looks roughly Gaussian. Comparing the two plots in Figure \ref{fig:resultsStep2} shows that the observational errors serve to widen the error ellipses somewhat in both directions.

The uncertainties we determine from the Step II KLD are more than an order of magnitude larger than the technical precision obtained by bootstrapping the data set and repeating Step I. This is because the bootstrap uncertainties represent the joint probability of the potential parameters and the action distribution produced by that potential. Since the action distribution depends on the number and properties of stream progenitors rather than the number of data points, resampling the data will always reconstruct the same action distribution, leading to very little change in the best-fit $\vect{a}$. Obtaining uncertainties on $\vect{a}$ independent of the action distribution via bootstrap would require resampling sets of streams instead of individual stars, using the stream-membership information we are trying to do without. Instead we use the Step II KLD to calculate the \emph{conditional} probability of $\vect{a}$ for the given action distribution, which gives reasonably sized uncertainties of roughly the same order as the distance between our best fit and the input value, as seen in Figure \ref{fig:resultsStep2}. 

The main weakness of this approach is that since we assume that the best-fit action distribution from Step I is the true distribution, the uncertainties only include values of $\vect{a}$ that produce distributions similar to the best fit, rather than any distribution with a similar Step I KLD. This is why some parameter combinations with high $\sub{\super{D}{I}}{KL}$ can have relatively low Step II probabilities: a good example is the case without observational errors in the left panel of Figure \ref{fig:resultsStep2}, where the second-best-fit (orange point) is less than half as probable as the best fit (red point) according to Step II. However, if we calculate $\sub{\super{D}{II}}{KL}$ with respect to the second-best-fit point, the resulting probability contours overlap with the ones pictured at better than the ``1-$\sigma$" level. The consistency between the size and shape of the $\sub{\super{D}{II}}{KL}$ contours, the spread and locations of the several highest values of $\sub{\super{D}{I}}{KL}$, and the difference between the best-fit and input values gives us confidence that this method of calculating the uncertainties gives reasonable values.

In the case with Gaia errors, the best-fit enclosed mass is $\sub{M}{enc,0}=0.208\substack{+0.085\\-0.043}\times10^{12} \Msun$ (corresponding to total mass $6.73\substack{+2.92\\-2.03}\times10^{12} \Msun$) and the best-fit scale radius is $11.97\substack{+2.57\\-3.03}$ kpc. The input value of the enclosed mass at the average distance of the stars in the sample (6.51 kpc, calculated using the error-convolved data), $\super{\sub{M}{enc}}{true}=0.215\times 10^{12}\Msun$, is well within the 68\% confidence interval---in fact, it is within 3\% of the true value. However the scale radius input value ($b=8$ kpc) is not within this interval, which means that the total mass $(M=2.7\times 10^{12}\Msun)$, isn't either since the two are linked through the enclosed mass. The error-free case behaves similarly though the confidence intervals are slightly smaller: $(\sub{M}{enc,0},b_0,M_0)=(0.223\substack{+0.082\\-0.038}\times 10^{12}\Msun,11.40\substack{+1.79\\-2.88}\textrm{ kpc},5.62\substack{+1.11\\-1.70}\times 10^{12}\Msun)$. The average distance is slightly larger when calculated without error convolution (6.85 kpc) so the input value for the enclosed mass in this case is slightly larger too: $\super{\sub{M}{enc}}{true}=0.240\times 10^{12}\Msun$. This value is within 7 percent of the best-fit (effectively, the grid point next to the one identified as the best fit in the case with errors). In short, the enclosed mass is recovered quite accurately in both cases while the recovered $b$ has larger relative error and is biased high in both cases, leading to errors in the total mass $M$. This is not an effect of the errors but a result of the relatively narrow range of galactocentric distances of the stars in the sample, shown in Figure \ref{fig:rdist}. A span of only about three scale radii, it appears, is not sufficient to eliminate this bias. We will show in Sections \ref{sec:varyNstream} and \ref{sec:dist} that adding stars that increase the distance range reduces the bias and improves the uncertainty on the scale radius and by extension the total mass.

\section{Performance with different numbers of streams}
\label{sec:varyNstream}

\begin{figure}
\begin{center}
\includegraphics[width=0.48\textwidth]{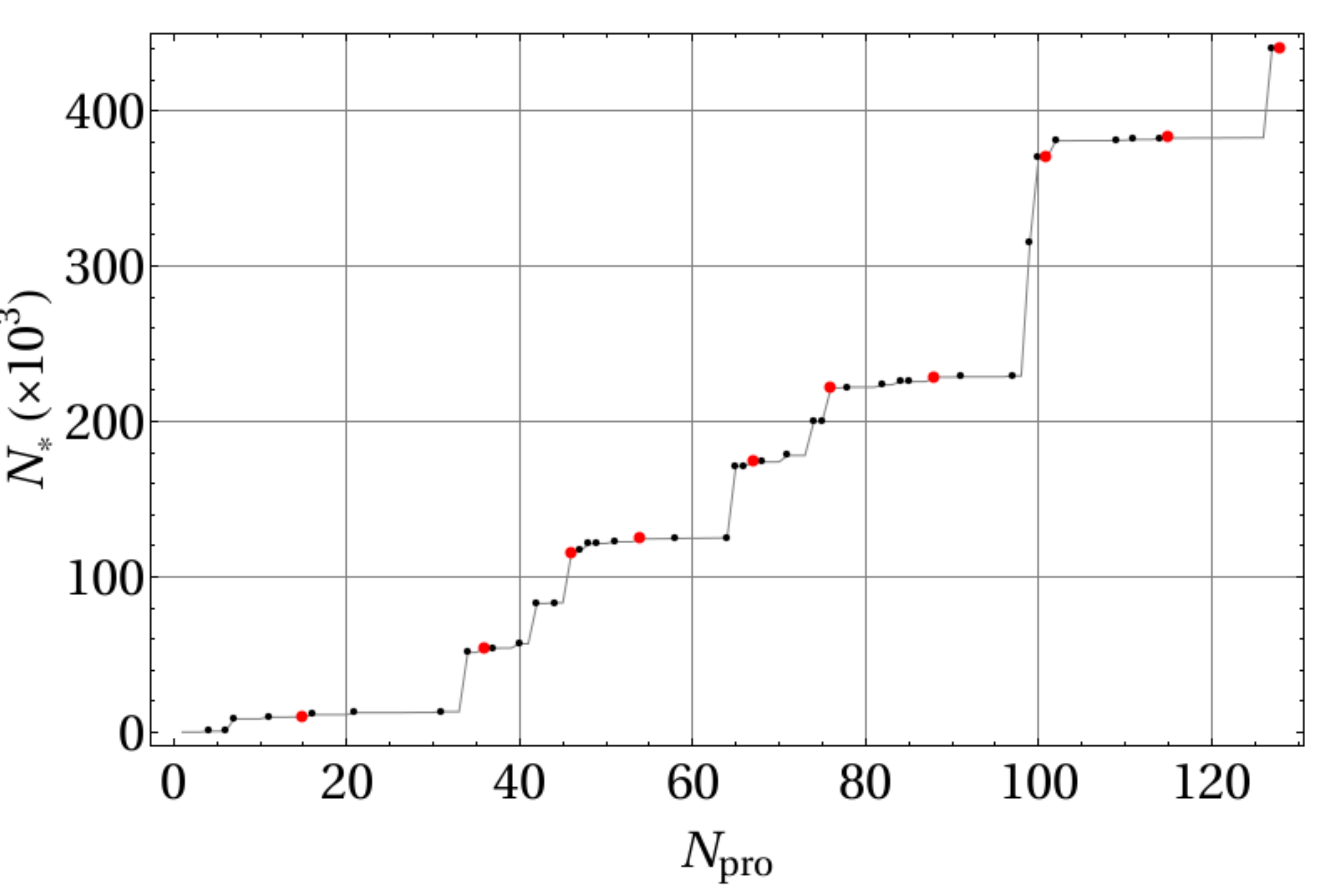}
\caption{The number of stars in subsequent subsamples of {\sc Gaia\_er} (or {\sc Gaia\_ne}), as a function of the number of different progenitors included in the sample. The $x$ axis here is the total number of progenitors (regardless of $n_*$), while the large black dots mark where an additional progenitor with more than 100 stars is added. The largest dots (red online), which are spaced by five $n_*>100$ streams, are the samples we analyzed using our fitting procedure.}
\label{fig:NstarPerStream}
\end{center}
\end{figure}

By selecting subsets of progenitors from the mock halo, we examined the sensitivity of the fit to the number and type of streams in a particular sample. We formed a random sequence of subsets of progenitors from {\sc Gaia\_er} by first selecting a single progenitor at random, then adding randomly-selected additional progenitors (without replacement) one by one. About half the progenitors in the sample have less than 100 stars present in {\sc Gaia\_er} (see the blue histogram in Figure \ref{fig:nstarstream}, and Table \ref{tbl:sampleStats}) and thus barely affect the fit at all, since our fit method weights each star equally. Therefore we only ``count'' progenitors with more than 100 stars in {\sc Gaia\_er} when building up the sample, though we do include the smaller clusters in the buildup sequence for consistency since they can function as noise. The result of this process, in terms of the number of stars in the samples as a function of the number of progenitors, is plotted in Figure \ref{fig:NstarPerStream}; the addition of each very large progenitor is apparent as a large jump in $N_*$. To determine the effect of the observational error we pulled the same sequence of subsamples from {\sc Gaia\_ne}.

\begin{figure}
\begin{center}
\includegraphics[width=0.48\textwidth]{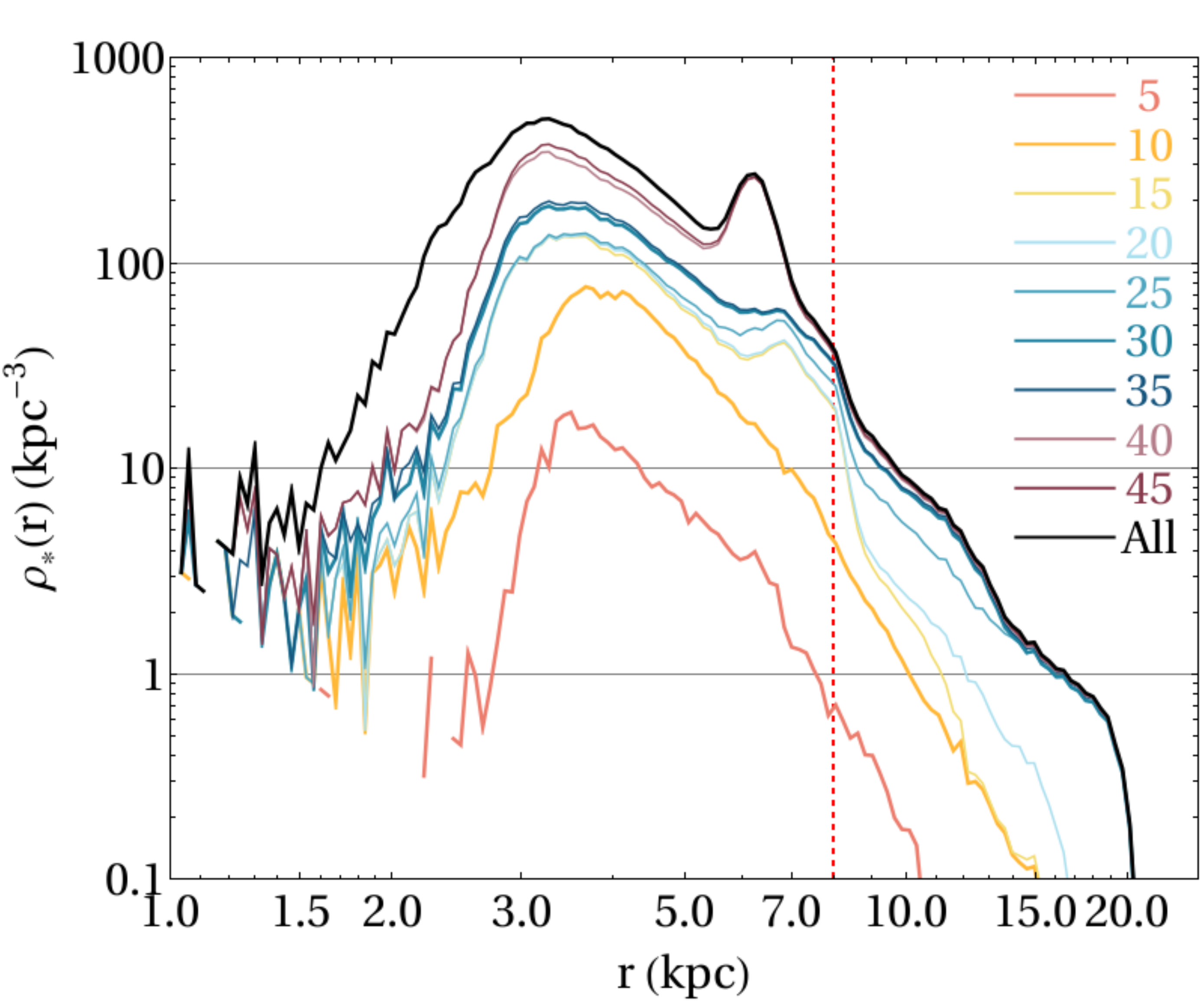}
\caption{Density distribution of galactocentric distances for stars in selected subsamples of {\sc Gaia\_er}, marked as red dots in Figure \ref{fig:NstarPerStream}. The black line is the distribution for the full sample. Lines in bold indicate the distributions for the subsamples whose fit results are plotted in Figure \ref{fig:VaryNStream}. As in Figure \ref{fig:rdist}, the red dashed line marks the scale radius of the input potential.}
\label{fig:subsampleRadiusDistribution}
\end{center}
\end{figure}

 Since the progenitors are added at random to build up the sample, we can also use these subsamples to study the relative influence of streams whose progenitors have orbits at different Galactocentric distances. The distance distribution of stars in various subsamples is plotted in Figure \ref{fig:subsampleRadiusDistribution}. In some cases a large progenitor enters at a particular distance; for example, the main difference between the $\sub{N}{pro}=10$ sample and the $\sub{N}{pro}=15$ sample, besides an overall increase in the number of stars, is the addition of a large progenitor with an orbit near the scale radius. On the other hand, the main difference between $\sub{N}{pro}=20$ and $\sub{N}{pro}=25$ is the addition of stars at larger radii.

\begin{figure*}
\begin{tabular}{ccc}
\includegraphics[width=0.3\textwidth]{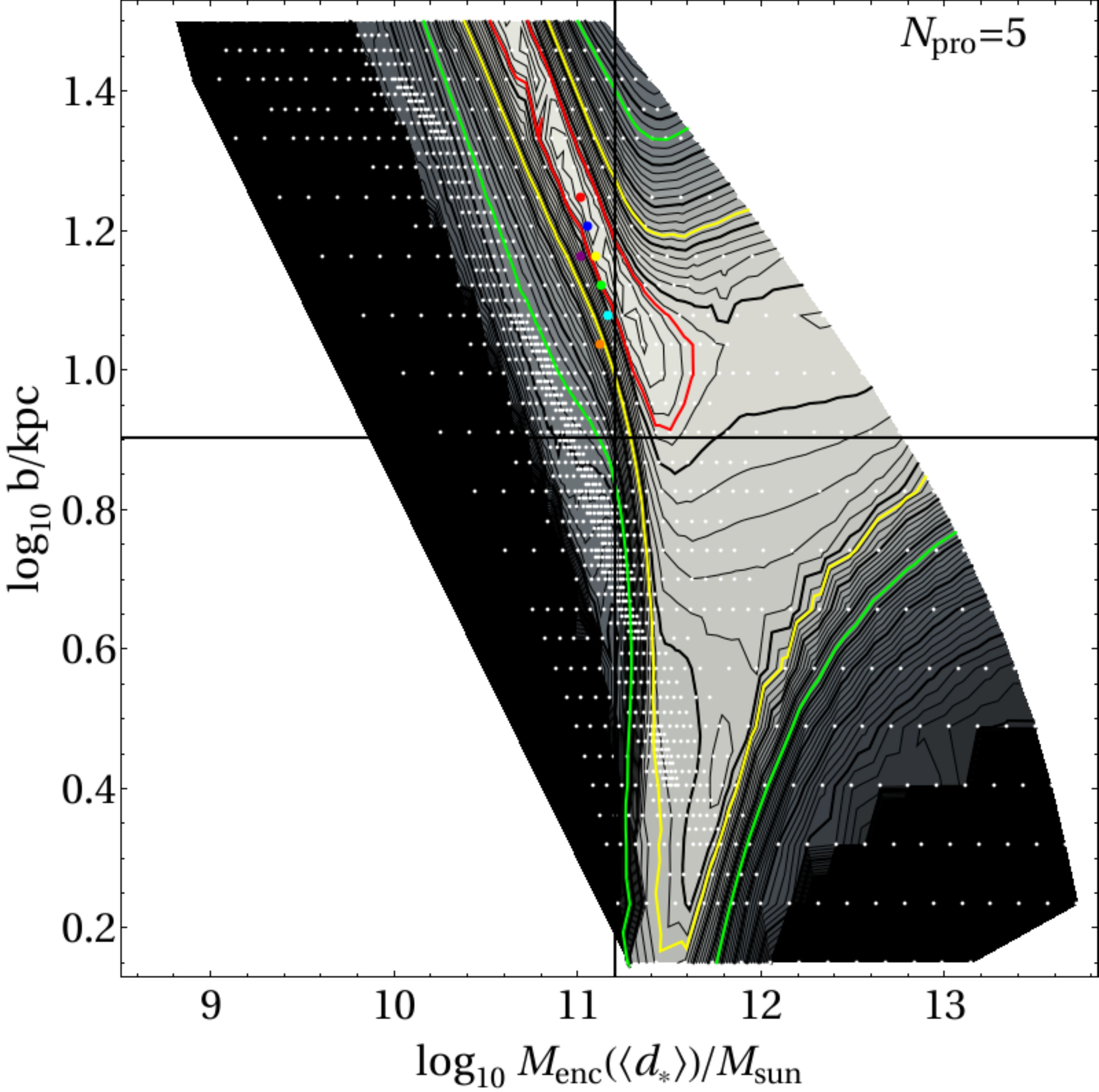} & \includegraphics[width=0.3\textwidth]{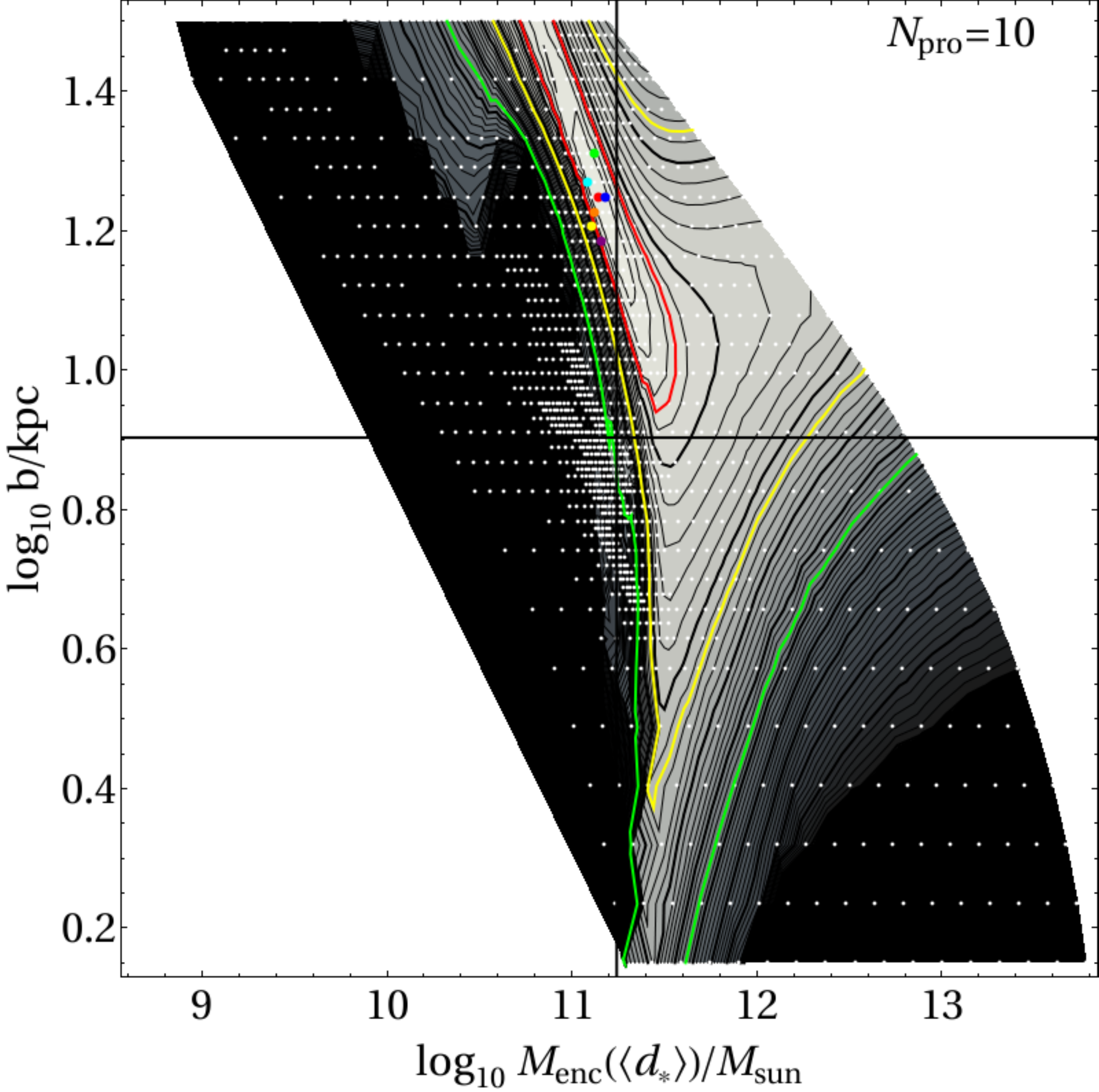} & \includegraphics[width=0.3\textwidth]{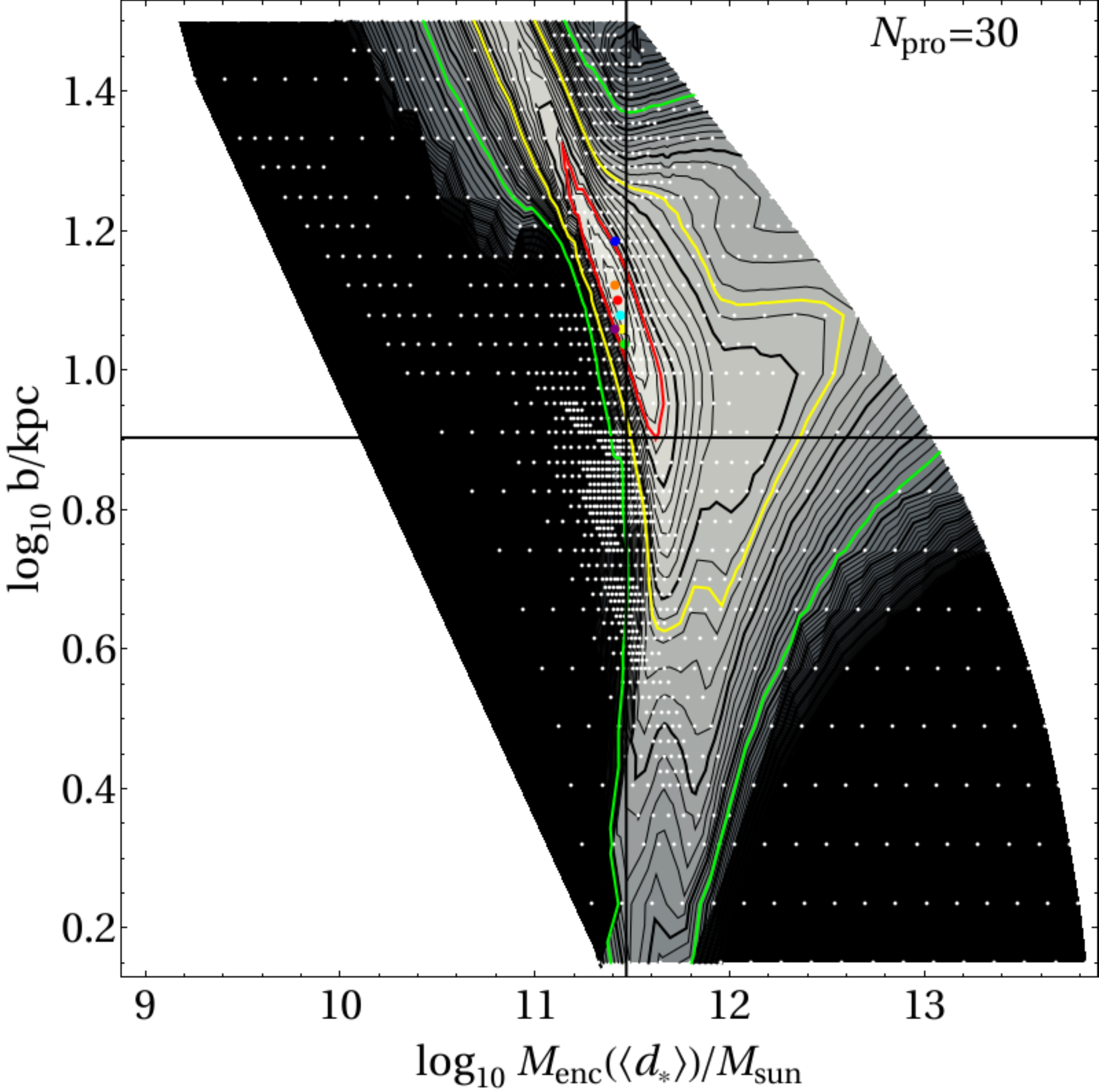}\\
\end{tabular}
\caption{As in Figure \ref{fig:resultsStep2}, but for streams from 5 (left), 10 (center), and 30 (right) progenitors with $n_* >100$ randomly selected from {\sc Gaia\_er}.}
\label{fig:VaryNStream}
\end{figure*}

\ifbool{SUBMIT}{
\begin{deluxetable}{cccccccccc}
\setlength{\tabcolsep}{0.02in}
\tabletypesize{\footnotesize}
\tablecaption{Fit results for {\sc Gaia\_er} \label{tbl:resultsEr}}
\tablewidth{0pt}
\tablehead{
\colhead{$\sub{N}{pro}$} & 
\colhead{$N_*$} & 
\colhead{$\langle d_* \rangle$} & 
\colhead{$\super{\sub{M}{true}}{enc}(\langle d_* \rangle)$} &
\colhead{$\super{M_0}{enc}$} &
\colhead{$M_0$ ($\sub{M}{true}=2.7$)} &
\colhead{$b_0$ ($\sub{b}{true}=8.0$)} &
\colhead{$\sub{\super{D}{I}}{KL}(\vect{a}_0)$} &
\colhead{$\super{\sub{D}{KL}}{I}(\sub{\vect{a}}{true})$} }
}{
\begin{table*}
 \caption{Fit results for {\sc Gaia\_er} \label{tbl:resultsEr}}

\begin{tabular}{ccccccccc}

\hline
$\sub{N}{pro}$ & 
$N_*$ & 
$\langle d_* \rangle$ & 
$\super{\sub{M}{true}}{enc}(\langle d_* \rangle)$ &
$\super{M_0}{enc}$ &
$M_0$ ($\sub{M}{true}=2.7$) &
$b_0$ ($\sub{b}{true}=8.0$) &
$\sub{\super{D}{I}}{KL}(\vect{a}_0)$ &
$\super{\sub{D}{KL}}{I}(\sub{\vect{a}}{true})$ \\
\hline
}

\ifbool{SUBMIT}{\startdata}{}

5 & \phantom{00}9 883 & 5.70 & 0.160 & $0.105\substack{\mathit{+0.185}\\-0.065}$ & $13.8\substack{\mathit{+20.2}\\-7.1}$ & $17.7\substack{\mathit{+14.0}\\-7.8}$ & 0.67 & 0.60  \\
10 & \phantom{0}53 966 & 5.92 & 0.175 & $0.140\substack{\mathit{+0.143}\\-0.076}$ & $16.5\substack{\mathit{+24.1}\\-9.8}$ & $17.7\substack{\mathit{+14.0}\\-7.8}$ & 0.90 & 0.68  \\
15 & 115 339 & 6.03 & 0.182 & $0.146\substack{\mathit{+0.129}\\-0.065}$ & $28.4\substack{\mathit{+29.9}\\-13.3}$ & $21.4\substack{\mathit{+10.2}\\-6.9}$ & 1.32 & 1.13  \\
20 & 124 459 & 6.32 & 0.202 & $0.183\substack{+0.119\\-0.088}$ & $13.8\substack{+14.6\\-5.0}$ & $16.0\substack{+10.0\\-4.0}$ & 1.49 & 1.34  \\
25 & 173 861 & 7.76 & 0.309 & $0.303\substack{+0.146\\-0.082}$ & $8.06\substack{+3.49\\-2.92}$ & $13.2\substack{+3.6\\-3.3}$ & 1.88 & 1.82  \\
30 & 221 570 & 7.58 & 0.295 & $0.269\substack{+0.127\\-0.086}$ & $6.73\substack{+2.92\\-2.03}$ & $12.6\substack{+4.2\\-2.7}$ & 1.88 & 1.80  \\
35 & 228 516 & 7.56 & 0.293 & $0.275\substack{+0.131\\-0.086}$ & $6.15\substack{+2.66\\-2.23}$ & $12.0\substack{+4.0\\-3.5}$ & 1.85 & 1.76  \\
40 & 369 954 & 6.87 & 0.242 & $0.236\substack{+0.095\\-0.051}$ & $9.65\substack{+4.18\\-3.49}$ & $13.8\substack{+3.0\\-3.5}$ & 2.09 & 2.02  \\
45 & 382 381 & 6.82 & 0.238 & $0.220\substack{+0.105\\-0.044}$ & $5.62\substack{+2.43\\-2.04}$ & $11.4\substack{+2.5\\-3.3}$ & 2.08 & 2.00  \\

All & 440 431& 6.51 & 0.215 & $0.208\substack{+0.085\\-0.043}$ & $6.73\substack{+2.92\\-2.03}$ & $12.0\substack{+2.6\\-3.0}$ & 2.03 & 1.97  

\ifbool{SUBMIT}{\enddata}{\end{tabular}}

\tablecomments{$\sub{N}{pro}$: Number of progenitors (with $n_*>100$) in sample. $N_*$: Total number of stars in sample.  $\langle d_* \rangle$: Mean galactocentric distance of stars in sample. $\super{\sub{M}{true}}{enc}(\langle d_* \rangle)$: Enclosed mass at mean radius in true potential. $\super{M_0}{enc}$: Enclosed mass at mean radius in best-fit potential. $M_0, b_0$: best-fit total mass and scale radius. $\sub{\super{D}{I}}{KL}(\vect{a}_0)$: Maximum value of the KLD found in Step I. $\super{\sub{D}{KL}}{I}(\sub{\vect{a}}{true})$: KLD of Step I (Equation \ref{eq:dkl1}) for the true values of the potential parameters. All masses are in units of $10^{12} M_\odot$; all distances are in kpc. Error bars indicate the extrema of the region where ($\super{\sub{D}{KL}}{II}\leq 1/2$); values in \emph{italics} denote an open contour (value at the boundary of the explored parameter space).}

\ifbool{SUBMIT}{\end{deluxetable}}{\end{table*}}

\ifbool{SUBMIT}{
\begin{deluxetable}{cccccccccc}
\setlength{\tabcolsep}{0.02in}
\tabletypesize{\footnotesize}
\tablecaption{Fit results for {\sc Gaia\_ne} \label{tbl:resultsNe}}
\tablewidth{0pt}
\tablehead{
\colhead{$\sub{N}{pro}$} & 
\colhead{$N_*$} & 
\colhead{$\langle d_* \rangle$} & 
\colhead{$\super{\sub{M}{true}}{enc}(\langle d_* \rangle)$} &
\colhead{$\super{M_0}{enc}$} &
\colhead{$M_0$ ($\sub{M}{true}=2.7$)} &
\colhead{$b_0$ ($\sub{b}{true}=8.0$)} &
\colhead{$\sub{\super{D}{I}}{KL}(\vect{a}_0)$} &
\colhead{$\super{\sub{D}{KL}}{I}(\sub{\vect{a}}{true})$} 
}
}{
\begin{table*}
 \caption{Fit results for {\sc Gaia\_ne} \label{tbl:resultsNe}}

\begin{tabular}{cccccccccc}

\hline
$\sub{N}{pro}$ & 
$N_*$ & 
$\langle d_* \rangle$ & 
$\super{\sub{M}{true}}{enc}(\langle d_* \rangle)$ &
$\super{M_0}{enc}$ &
$M_0$ ($\sub{M}{true}=2.7$) &
$b_0$ ($\sub{b}{true}=8.0$) &
$\sub{\super{D}{I}}{KL}(\vect{a}_0)$ &
$\super{\sub{D}{KL}}{I}(\sub{\vect{a}}{true})$\\
\hline
}

\ifbool{SUBMIT}{\startdata}{\null}

5 &  \phantom{00}9 883 & 5.79 & 0.166 & $0.138\substack{+0.112\\-0.085}$ & $4.70\substack{+6.85\\-1.96}$ & $10.9\substack{+10.6\\-4.2}$ & 0.67 & 0.64  \\
10 &  \phantom{0}53 966 & 6.23 & 0.196 & $0.198\substack{+0.112\\-0.096}$ & $5.62\substack{+8.20\\-2.63}$ & $10.9\substack{+8.6\\-3.9}$ & 1.07 & 0.90  \\
15 & 115 339 & 6.37 & 0.205 & $0.192\substack{+0.098\\-0.109}$ & $5.14\substack{+8.68\\-1.86}$ & $10.9\substack{+10.6\\-3.1}$ & 1.51 & 1.39  \\
20 & 124 459 & 6.70 & 0.229 & $0.229\substack{+0.070\\-0.069}$ & $4.29\substack{+1.86\\-1.02}$ & $\phantom{1}9.86\substack{+3.33\\-2.13}$ & 1.83 & 1.78  \\
25 & 173 861 & 8.33 & 0.354 & $0.348\substack{+0.069\\-0.065}$ & $3.00\substack{+0.28\\-0.49}$ & $\phantom{1}8.52\substack{+1.34\\-1.50}$ & 2.37 & 2.34  \\
30 & 221 570 & 8.13 & 0.338 & $0.330\substack{+0.070\\-0.048}$ & $3.28\substack{+0.47\\-0.54}$ & $\phantom{1}8.94\substack{+0.91\\-1.58}$ & 2.35 & 2.31 \\
35 & 228 516 & 8.12 & 0.337 & $0.330\substack{+0.068\\-0.036}$ & $3.13\substack{+0.45\\-0.63}$ & $\phantom{1}8.73\substack{+1.13\\-1.71}$ & 2.35 & 2.30  \\
40 & 369 954 & 7.26 & 0.271 & $0.254\substack{+0.061\\-0.046}$ & $4.70\substack{+0.93\\-1.11}$ & $10.6\substack{+1.37\\-2.08}$ & 2.53 & 2.45  \\
45 & 382 381 & 7.21 & 0.266 & $0.270\substack{+0.044\\-0.040}$ & $4.29\substack{+0.85\\-1.02}$ & $\phantom{1}9.86\substack{+1.55\\-1.74}$ & 2.50 & 2.43 \\

All & 440 431& 6.85 & 0.240 & $ 0.223\substack{+0.082\\-0.038}$ & $5.62\substack{+1.11\\-1.70}$ & $11.40\substack{+1.79\\-2.88}$ & 2.41 & 2.34  \\

\ifbool{SUBMIT}{\enddata}{\end{tabular}}

\tablecomments{Column headings, units and error bars are the same as in Table \ref{tbl:resultsEr}.}

\ifbool{SUBMIT}{\end{deluxetable}}{\end{table*}}

Figure \ref{fig:VaryNStream} shows the results from a few of the subsamples of {\sc Gaia\_er} that illustrate the effect of adding large streams at various galactocentric distances. The leftmost panel shows the result with only five progenitors of more than 100 stars, including streams from only one moderately large progenitor of several thousand stars. The stars in this subsample span less than one scale radius in galactocentric distance, between about 3 and 10 kpc, and are mostly inside the scale radius (see the thick dotted red line in Figure \ref{fig:subsampleRadiusDistribution}). Despite the small number of unique progenitors, the enclosed mass measured by this sample is already within 35 percent of the input value, though the half-probability contour is not yet closed at large $b$. However, because of the small range of distances explored by the orbits of the progenitors, the scale radius is not recovered (the error contours are not closed) and the best-fit value is fairly inaccurate, not even within a factor of 2.

Adding five more progenitors, including a very large one, improves the accuracy with which the enclosed mass can be recovered, as shown in the center panel of Figure \ref{fig:VaryNStream}. This sample has roughly the same distance distribution of stars (shown as the thick yellow line in Figure \ref{fig:subsampleRadiusDistribution}) as the 5-progenitor sample, just with more stars at every distance---about five times as many stars in total---and a slightly larger distance range, about 1.5 scale radii. Thus this sample is no better at measuring the scale radius, but the confidence contours have gotten slightly smaller, and now the measured value of the enclosed mass is within 20 percent of the input value. We can also see, by comparing to the five-progenitor sample, that simply adding more stars is not the most effective way to improve the measurement.

Adding progenitors with orbits at still larger radii to increase the distance range again, however, does start to improve the scale radius measurement, as shown in the right panel of Figure \ref{fig:VaryNStream}. This sample contains streams from a total of 30 progenitors, including 6 of the 9 largest, and has roughly the same distance range as the entire sample (about 2.5 scale radii), including all the stars at the largest distances. This is sufficient to close the half-probability contour. The best-fit value of $b$ has less bias, as well: it is now within about 60 percent of the input value. The best-fit enclosed mass is within 9 percent.

Tables \ref{tbl:resultsEr} and \ref{tbl:resultsNe} summarize our results in steps of five progenitors each, up to the full sample. The tables include the number of progenitors and stars in a given sample, the average distance and enclosed mass in each case, and the recovered parameters with error bars indicating the extent of the half-probability contour ($\super{\sub{D}{KL}}{II}<1/2$). The last two columns give a sense of the difference in KLD values for Step I between the true and best-fit potentials. 

\begin{figure*}
 \plottwo{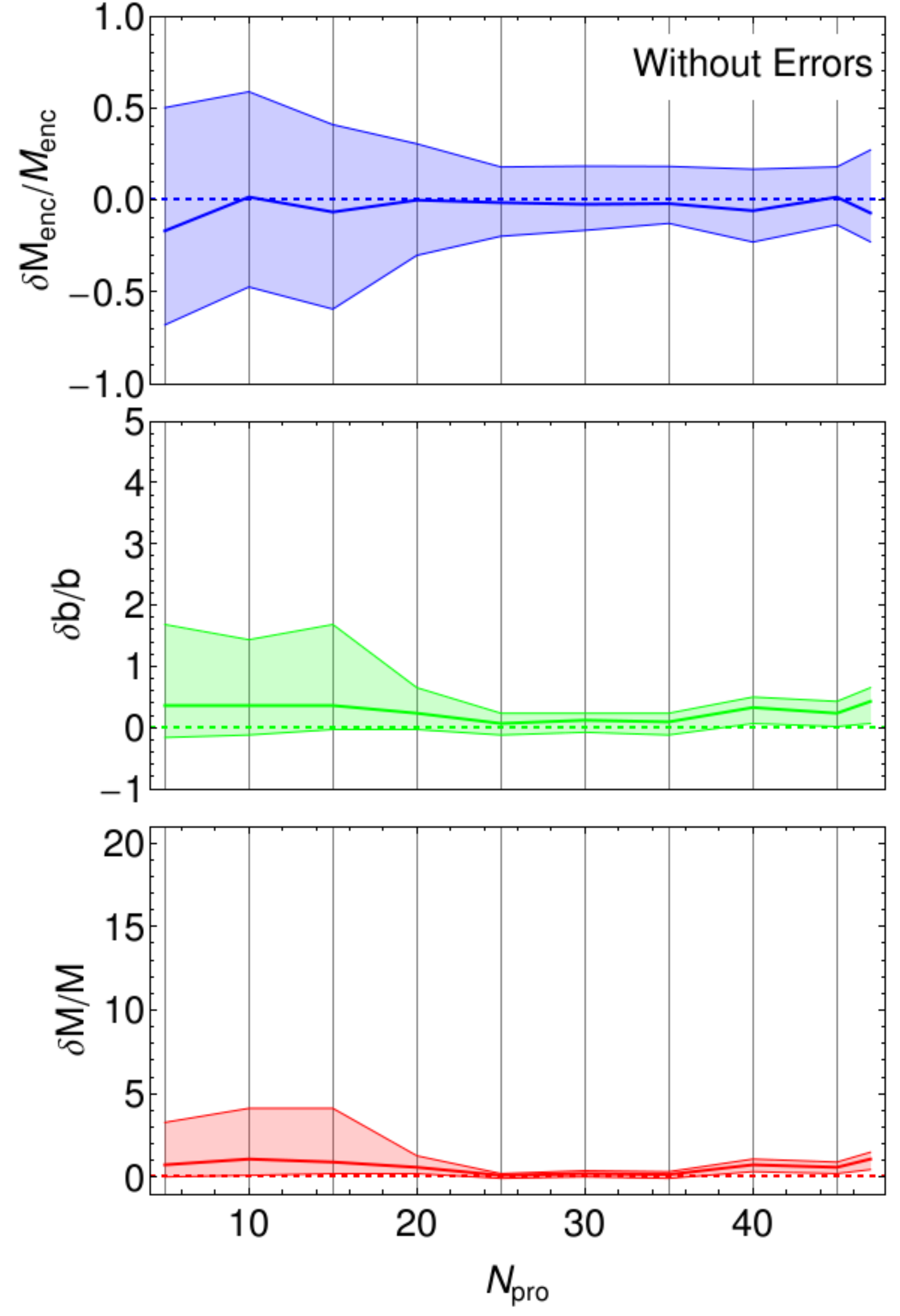}{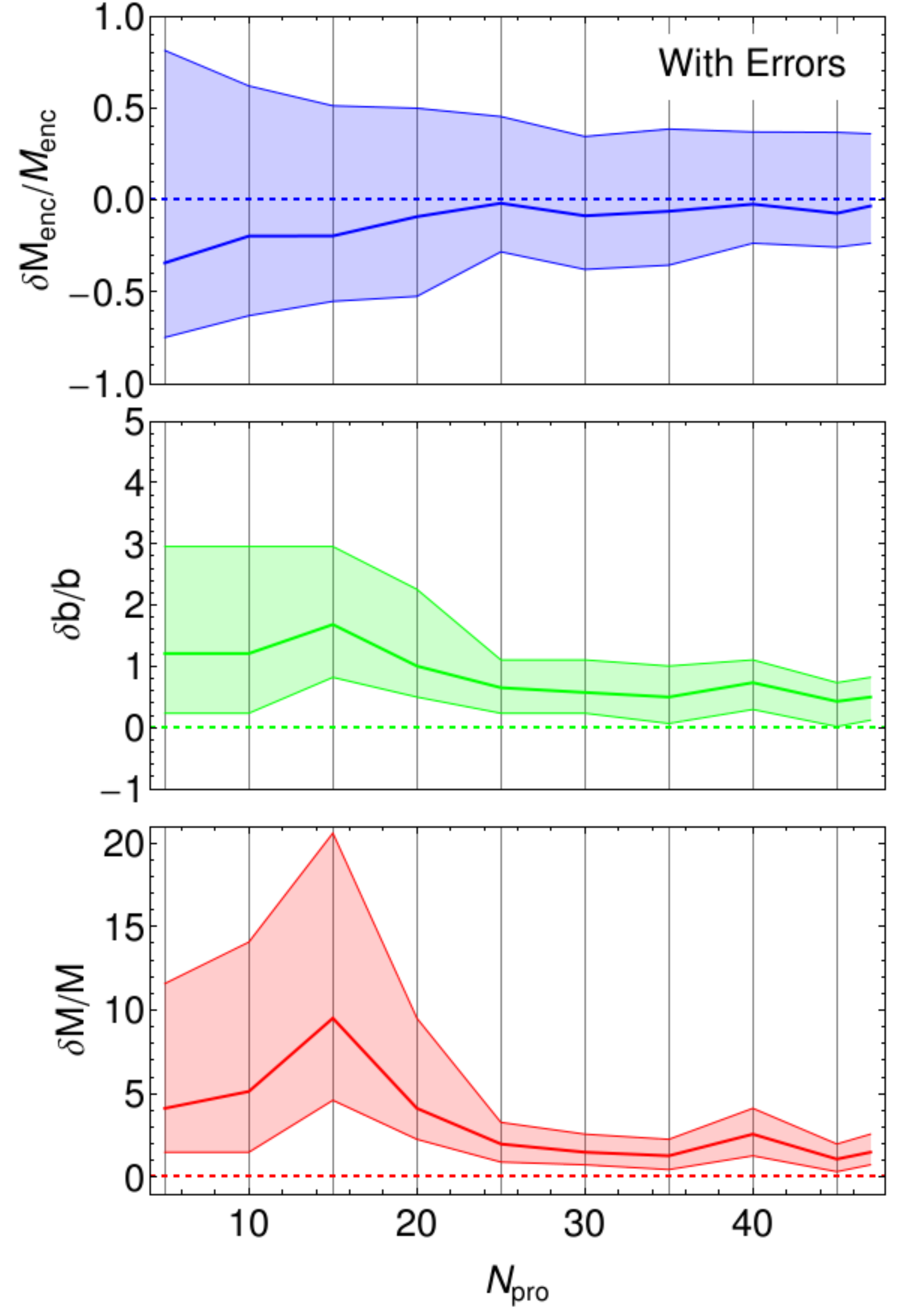}
\caption{Relative difference, $\delta x/x \equiv (x_0-\sub{x}{true})/\sub{x}{true}$, between the best-fit and input value for the enclosed mass $\sub{M}{enc}$ (top; blue), scale radius $b$ (middle, green), and total mass $M$ (bottom, red). The central line shows the best-fit; the shaded regions indicate the error bars derived from the $\super{\sub{D}{KL}}{II}=1/2$ relative probability contour. The left panel shows results for subsamples of {\sc Gaia\_ne}, the right panel for {\sc Gaia\_er}. }
\label{fig:errorBehavior}
\end{figure*}

In Figure \ref{fig:errorBehavior} we plot how well the algorithm recovers the input values of the enclosed mass, scale radius, and total mass as a function of the number of unique progenitors (with $n_*>100$) in the sample. Note that the y-axes of the plots for each parameter are significantly different: the enclosed mass is by far the most accurately and precisely determined, followed by scale radius and then total mass. Comparing the left panel, which shows the results for subsamples of {\sc Gaia\_ne}, with the right panel, which shows {\sc Gaia\_er} subsamples, shows that the observational errors increase the error bars on all three quantities: by a factor of $\sim$ 1.5 for the enclosed mass, $\sim$2 for the scale radius, and $\sim$2--4 for the total mass, which is the least accurately determined. Regardless of observational error, about 20-25 progenitors are enough to get the error bar sizes to converge, though error does affect how many streams are necessary to bound all the fit parameters from above and below: for fewer than 15 progenitors the confidence intervals extend beyond the bounds of our explored parameter space when errors are included, yet are closed for error-free data.. The observational errors seem to be responsible for larger differences between the best-fit and input parameters mainly for small numbers of progenitors, though the bias in $b$ (and hence $M$) is slightly larger for the error-convolved sample even for many progenitors. However, the bias in $b$ is definitely present even in the error-free sample, which indicates that the streams making up the sample are themselves responsible for at least part of it.

Examining the tables shows a few more trends. We see that when including observational errors, streams from about 15 progenitors with $n_*>100$ (including two or three large ones) are needed to close the $\super{\sub{D}{KL}}{II}=1/2$ relative probability contour. Comparing the results with and without errors shows that the effect of the observational errors is to require more streams to get a bounded measurement at the $\super{\sub{D}{KL}}{II}=1/2$ relative probability level. We also see that the Step I KLD values are generally higher for the samples without errors, since the errors tend to increase the size of the individual action space clumps. In both samples, with streams from up to about 30 or 40 progenitors the KLD of Step I increases steadily as more are added, reflecting the increasing amount of action-space information. The probabilities of the best-fit point and the input parameters also steadily converge in this regime. However, one can also see from these trends that the addition of streams from the largest progenitors is the most influential: for example, between 25 and 35 progenitors only one medium-sized progenitor is added and the overall radial distribution stays roughly the same (Figure \ref{fig:subsampleRadiusDistribution}), so there is not much difference in the fit results. Interestingly, the 45-progenitor sample appears to do somewhat better than the full sample; from Figures \ref{fig:NstarPerStream} and \ref{fig:subsampleRadiusDistribution} we see that the difference is one large stream at small radius. This shows how crowding and overlap of structures deep in the potential can decrease the information and degrade the fit, and suggests that perhaps a slightly different energy selection could improve the results.

\section{The effect of increasing the distance range}
\label{sec:dist}

The results from fitting different numbers of progenitors suggest that a larger range of distances in the sample will alleviate the bias in determining the scale radius with our method. The distances of stars in our mock halo are limited mainly by the requirement that all six phase-space coordinates be measured by Gaia. Since proper motions are measured to $V \approx 20$, but radial velocities only to $V \approx 17$, there are many distant stars in the Gaia data set that will be missing this final coordinate. The planned spectroscopic surveys WEAVE \citep{Dalton2012} and 4MOST \citep{DeJong2012} aim to obtain the missing radial velocities for at least some of these faint halo stars. 

We can estimate the effect of completing the six-dimensional catalog by augmenting the sample {\sc Gaia\_ne}\footnote{To keep the comparison direct and simple, we will forgo incorporating error convolution and just deal with the unconvolved positions and velocities, since RVs from ground-based follow-up will have different error properties than the ones provided by Gaia.} with stars that would be included in the sample but lack radial velocity measurements from Gaia: that is, any star in our mock halo with $V\geq 17.3$. Adding these stars (61,893 of them) to the sample extends the distance range to nearly 100 kpc from the Galactic center and increases the average distance from 6.85 kpc (inside the scale radius) to 8.60 kpc (slightly outside the scale radius). Beyond the scale radius the radial distribution of this new sample roughly follows that of the full mock halo (black line in Figure \ref{fig:rdist}). 

\begin{figure}
 \plotone{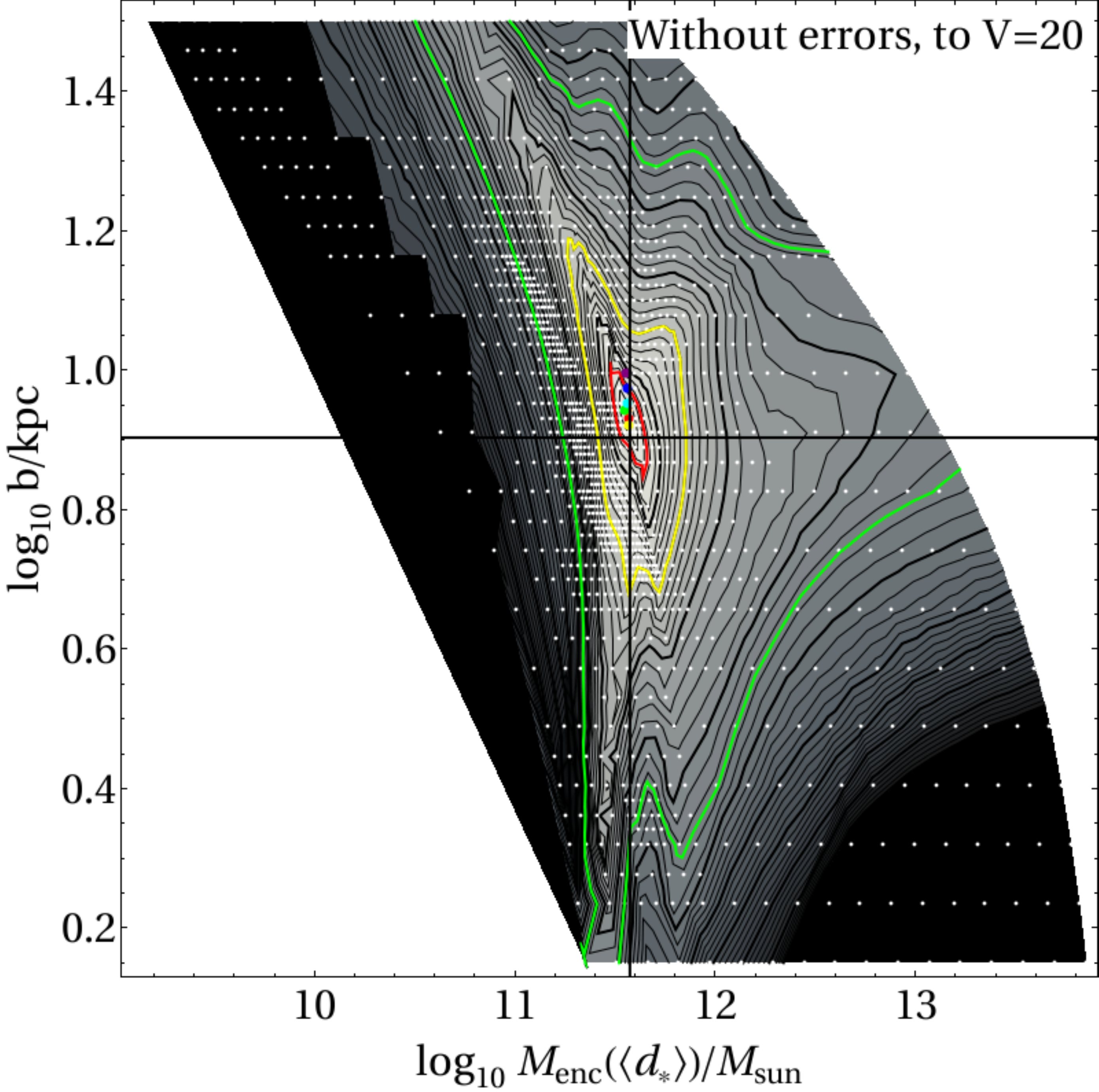}
\caption{As in Figure \ref{fig:resultsStep2}, but for a sample without errors including stars to $V=20$ (approx. 100 kpc).}
\label{fig:step2toV20}
\end{figure}

 Figure \ref{fig:step2toV20} shows the effect on the confidence contours of increasing the distance range. As predicted, the bias in the $b$ measurement is nearly eliminated. Thanks to the addition of a few extra distant progenitors (53 with $n_* > 100$ as opposed to 47) as well as the improved location along the enclosed-mass degeneracy, the error contours are also slightly smaller than for {\sc Gaia\_ne} alone (compare the left panel of Figure \ref{fig:resultsStep2}). With this new sample, we obtain the best-fit total mass $M_0 = 3.00\substack{+0.59\\-0.49} \times 10^{12}\ M_{\odot}$, the best-fit enclosed mass $\sub{M}{enc,0} = 0.370\substack{+0.073\\-0.067} \times 10^{12}\ M_{\odot}$ (true value: $0.375\times 10^{12}\ M_{\odot}$), and the best-fit scale radius $b = 8.52\substack{+1.34\\-1.16}$ kpc. In other words, with this sample the recovered total mass is within 11 percent of the true value and has $\lesssim 20$\% uncertainties, the recovered enclosed mass is within 1.3 percent of the true value with $\lesssim 20$\% uncertainties, and the recovered scale radius is within 6.5 percent of the true value with $\lesssim 16$\% uncertainties. The measurement of the scale radius is still biased slightly high, but the input value, like that of the total mass, is now within the confidence interval of the recovered result.

\section{Discussion \& Conclusions}
\label{sec:concl}

The strategy of fitting a potential model using action clustering, as presented in this work, has several important advantages. Most importantly, it correctly recovers the input parameters in our simple example as long as the stars in the sample span a sufficient range in distance. Second, this method does \emph{not} require assigning stars to a particular stream, nor that all the stars in the sample belong to a stream, as long as the action-space distribution is dominated by well-separated structures. In this work we verify this by eye, by guessing a potential, which does not need to be too close to the real one, and looking at the distribution in the energy-angular momentum space.  The streams to be fitted occupy the portion of this space that looks clumpy to the eye, as described in Section \ref{subsec:streamsel}. Third, the fitted enclosed mass tends to be more accurate than it is precise; that is, the recovered enclosed mass is generally much closer to the input values than the limits of the confidence interval. The bias in the scale radius in our results is due to the small number of scale radii spanned by the stars in our sample, since adding stars that increase the distance range reduces this bias, as discussed in Sections \ref{sec:varyNstream} and \ref{sec:dist}. Finally, use of the Kullback-Leibler divergence as the figure of merit in the fit gives a well-defined maximum and an estimate of the confidence region for the potential parameters, thanks to its interpretation as a relative probability. The KLD is positive-definite and fast to compute using the method we describe in Section \ref{subsec:klcomp}, which bodes well for exploring higher-dimensional parameter spaces. 

Although we emphasize here that this method does not require the identification of individual streams, stream membership information can improve the method by allowing us to change the relative weighting of the stars in various streams. The version of our method we show here is completely blind to stream membership, weighing equally the contribution of each star to the KLD. This means that the largest clumps have the greatest influence on the result simply because they contain orders of magnitude more stars than the smallest structures. However, in reality the self-gravity of the progenitor satellite couples the number of stars to their extent in action space, as both increase with the mass of the progenitor. In fact the thinnest streams, which contain the fewest stars, are the most sensitive to the potential because they form the smallest and tightest action-space clumps; i.e., they are the most informative. The ability to either choose only the thinnest streams, or to more heavily weight their contribution relative to the larger, less-informative streams, should greatly improve the fit results.

\subsection{Analysis of results}

Our results suggest that our method's ability to measure the Galactic scale radius depends on the steepness of the radial density profile of the stellar halo. In our tests we used a halo that is quite centrally concentrated, both by adopting a fairly steep radial profile ($\rho \propto r^{-3.5}$) and by placing the few largest satellites on small-apocenter orbits. If the stellar halo is really this centrally concentrated, then it will be quite difficult to obtain a distance range of more than a few scale radii, given that the dark halo scale radius is now estimated to be roughly 10--20 kpc \citep{Battaglia2006,Battaglia2005,Navarro2010,Deason2012,Kafle2012,Sofue2012,Nesti2013}, or even larger depending on the model assumed \citep{Irrgang2013}---at any rate, it may be significantly larger than the 8 kpc used in our tests. However, a log-slope of $-3.5$ may be too steep to model the Galactic halo as a single power law; a more likely value is $-2.5$ \citep{DePropris2010} to $-2.9$ \citep{Deason2011}. Broken power laws provide a better fit, usually with an inner log-slope of $-2.3$ to $-2.8$ out to 30--45 kpc \citep{Bell2008,Deason2011,Sesar2011,Akhter2012} with a steeper fall-off (log-slope $-3.8$ to $-5$) at larger radii. All these measurements point toward more stars at larger radius relative to our toy model, which should improve our ability to measure the scale radius relative to what is presented here.

The results of this work have a few important implications for observations. First, the data obtained by Gaia will indeed be good enough to simultaneously fit multiple streams using this method: as seen in Figure \ref{fig:resultsStep2}, the observational errors widen the confidence contours only a little. Second, adding more stars of similar data quality, for example by obtaining radial velocities for faint stars via ground-based spectroscopic follow-up \citep{Dalton2012,DeJong2012}, will improve our ability to constrain the potential parameters by bringing in stars at larger distances that in the Gaia catalog alone are missing RVs. Finally, we find that for our mock halo, this technique requires  streams from $\sim$15 progenitors in order to provide both upper and lower bounds on the fit parameters and $\sim$20-25 to stabilize the uncertainties, well below the number expected to exist in the Milky Way; this is a good sign since galactic potentials with more parameters will likely require a higher number of streams. 

\subsection{Comparison with other stream-fitting methods}

Several other methods have been developed for measuring the potential by fitting orbits to streams; however, to our knowledge the method presented here is the only one that does not require assigning membership of stars to streams. Stream-fitting methods tend to fall roughly into two classes: those that fit an orbit or simulated stream to the stars' positions in phase space, such as the recent work by \citet{2014arXiv1406.2243G}, \citet{Price-Whelan2013}, and \citet{2013ApJ...773L...4V}, and those that exploit clustering or correlations in the space of actions (or other constants of motion like energy), angles, and frequencies \citep{2012ApJ...760....2P,2013MNRAS.433.1826S}. 

Methods that fit orbits or simulated streams usually define the goodness of fit (and hence the uncertainties on the fitted parameters) via distance between the model and observations in the observed coordinates, \`a la \citet{2010ApJ...714..229L}. The difficulty of obtaining the uncertainties for these methods is in having satisfactorily explored a sufficient percentage of a high-dimensional parameter space (one must usually explore the initial conditions and properties for the disrupting satellite as well as the parameters of the potential) with a likelihood function that is costly to evaluate (for each combination of parameters tried, a new simulated stream must be created). The assignment of membership in the stream to particular stars at key points along the stream can also affect the results. The most recent measurement of the total mass from the Sagittarius stream, by \citet{2014arXiv1406.2243G}, quotes 10\% uncertainties on the total mass at 100 kpc (the distance to the farthest stars) assuming a specific mass profile for the halo. This is better than the performance of our method (20\% uncertainties) given data to the same distance, but relies on knowledge of the progenitor of the stream to reduce significantly the number of fit parameters. Most known streams cannot be connected with their progenitors.

Methods that use clustering in the space of constants of motion, on the other hand, avoid having quite as large a parameter space but also usually lack a straightforward way of determinining uncertainties. Two examples are the fitting algorithm by \citet{2013MNRAS.433.1826S}, which uses correlations in angle-frequency space, and the one by \citet{2012ApJ...760....2P}, which uses the clustering in energy. So far, both of these methods still require identifying stream membership and have been demonstrated on single streams only. Both methods will also have a bias for time-dependent potentials, where the angles are no longer strictly proportional to their corresponding present-day frequencies (as assumed by \citeauthor{2013MNRAS.433.1826S}) and energy is not conserved (leading to less sensitivity in \citeauthor{2012ApJ...760....2P}'s method). 

The method closest to ours is the one proposed in \citet{2012ApJ...760....2P}. This method obtains a fit by minimizing the \emph{absolute} entropy of the energy distribution instead of the \emph{relative} entropy (KLD) of the action distribution. Indeed, the energy-angular momentum distribution of streams also becomes less clumpy for incorrect potentials. Energies also have the significant advantage of being much easier to calculate than the actions for realistic mass models.  However, the absolute entropy lacks the direct relationship to likelihood ratios that allows us to calculate confidence intervals from the KLD. One could substitute the KLD for the absolute entropy, but in this case the energy distribution has other features that make it more difficult to use than the actions. One is the sharp edge of the distribution as seen in Figure \ref{fig:trueActions}, which will vary in shape as the parameters change and complicate the calculation of the KLD between neighboring distributions in parameter space. Also, unlike the radial action, energy clumps not only scale in size as the parameters vary (the effect we scaled $J_r$ to eliminate), but the overall offset of the distribution also changes. Thus comparing the best-fit distribution to trial distributions in Step II of the procedure would require an adjustable offset chosen to realign the two distributions for each combination of parameters. Although the results of Step I show better contrast in our tests with energy replacing $J_r$, we expect this advantage to disappear for time-dependent potentials where the action is the adiabatic invariant, not the energy. Finally, in our spherical example it is straightforward to replace $J_r$ with $E$ since they are both conserved quantities and the only one of the three dimensions to depend on the potential parameters, but for nonspherical potentials the question is which action(s) one would replace with an energy: axisymmetric potentials, for example, have two potential-dependent actions that could be reinterpreted in terms of an energy. For these reasons, we use the action distribution in this work.

The method presented in this work uses only three of the six possible phase space coordinates to carry out the fit; we discard the angles, which for a fully phase-mixed stream will be evenly distributed on $[0,2\pi)$ but are indeed correlated with the actions and/or frequencies in potentials close to the correct one, as is exploited by \citeauthor{2013MNRAS.433.1826S} in their fitting method. Since all the streams in a sample, regardless of their degree of phase mixing, will have angles that fall in the same range, including this information in our fitting procedure would only be possible if stream membership information was also included. The price to avoid assigning membership in our method is that we throw out the angle information that is retained in the \citeauthor{2013MNRAS.433.1826S} method and in the entire class of orbit-fitting methods. This may be the reason for the extreme sensitivity of the \citeauthor{2013MNRAS.433.1826S} method, for example. It may also be a reason to prefer six-dimensional methods over our three-dimensional one if good stream membership information is available: they are potentially more powerful since they include phase information.

\subsection{Caveats}

This method contains two sources of bias that are not addressed in our toy model. The first is the possibility that substructure within individual action-space clumps can cause a bias, by preferring potentials that force these substructures to overlap into one clump. We do expect to see at least bimodal substructure in real tidal streams, which lose stars primarily from their two Lagrange points during tidal stripping. This selection effect produces a pair of subclumps for each stream when the self-gravity of its progenitor galaxy is properly accounted for. We integrate our progenitors as test particles, so this effect is not present in our tests, but we anticipate that the choice of potential that would unify each pair of subclumps will be different for each stream, and will cancel out when more than a few streams are present. 

A second possible source of bias is related to the fact that streams do not exactly follow a single orbit, and that their stars are sorted by energy along the stream. This means that the stars' actions (or frequencies) and angles are correlated, an effect which is exploited by \citet{2013MNRAS.433.1826S} in their fitting method. Computing actions with an incorrect potential in some sense mixes the true actions and angles of the stream stars, and one could in principle find a potential that does this in such a way to exactly cancel the action-angle correlation, producing a clump in action-space that is actually smaller than for the true potential. However, as in the case of clump substructure, we think it likely that the potential that would do this will be different for every stream, so that the bias cancels out for samples containing more than a few streams.

\subsection{Future work}
An outstanding question from this work is how the fit performance will be affected when galactic potentials with fewer symmetries are used. Generally, including more parameters in a fit to the same data tends to degrade fit performance, but this is not obviously the case for our method. In the spherically-symmetric toy models used in this work, only one of the three actions, $J_r$, is sensitive to the gravitational potential. This means that the change is the KLD is driven entirely by changes in only the $J_r$ distribution. For axisymmetric potentials, two of the three actions depend on the potential; in a triaxial potential all three will change as the potential is adjusted. This means that for less symmetric cases the change in the KLD as one moves away from the correct potential should be larger than for the spherically-symmetric case, which would imply that the fit will perform \emph{better} for less symmetric potentials. On the other hand, there is no formal proof that all the actions will be maximally clustered for the same choice of trial potential; it could be that a slightly different potential produces the most clustered actions in each dimension, which could again degrade fit performance. Determining which of these two possibilities dominates the fit's behavior will be the subject of a subsequent paper.

Although the method we propose does not require stream membership information, it can be extended to incorporate this information if it is available, most naturally by including metallicity or chemical abundance information as extra dimensions. Each additional clustering dimension improves the contrast between the best fit and other potentials, even if that dimension itself is not sensitive to the potential. This is why we include all three actions in our spherically symmetric toy model fits, even though only $J_r$ changes with the potential parameters; likewise, we expect stream stars to have some clustering in their chemical properties as well \citep{2002ARA&A..40..487F,2009ARA&A..47..371T}. The main technical issues in incorporating chemical information are extending the density estimation and KLD calculation machinery to a sufficient number of dimensions, and determining the appropriate relative weighting of the action and chemical subspaces (i.e., establishing the metric for computing distances). In future work we plan to explore this possibility further.

Our method is limited by the requirement that six-dimensional positions be available for all stars in the sample, and further by our naive treatment of individual stars as point measurements instead of error distributions. Including stars with missing coordinates would allow us to exploit a far wider range of kinematic data. Using points to represent the stars also forces us to throw out stars whose distance errors result in an unbound orbit, although these contain information, and does not correctly propagate the observational errors through into action space (a transformation that can produce a multi-peaked error distribution in the actions from Gaussian errors in observed space). Both these problems can be solved by working with probability distributions instead of point measurements, an innovation that we hope to incorporate in future work.

This method provides a framework for exploring various important questions related to potential constraints. First, the real Milky Way potential is not an exact copy of a parameterized model. We intend to explore the effect of imposing a fitted parameterization different from the input model: what parameters are recovered and what is their relation to the true potential? Second, some of the most interesting features of the halo concern its departures from spherical symmetry; we intend to expand the method to handle at least axisymmetric halos so that we can determine how well streams will constrain the flattening. Third, actions are superior to generic constants of the motion (such as energy) in part because they are adiabatically invariant; we intend to explore the effect of a slowly growing Milky Way in future work. Finally, this work considered only a single stellar luminosity, whereas in reality tidal streams have populations of stars. Using a more realistic model for the stream stars' types can produce better forecasts for the power of Gaia and other surveys to discover the gravitational potential of our Galaxy.

\begin{figure}
\ifbool{SUBMIT}{\epsscale{0.55}}{}
 \plotone{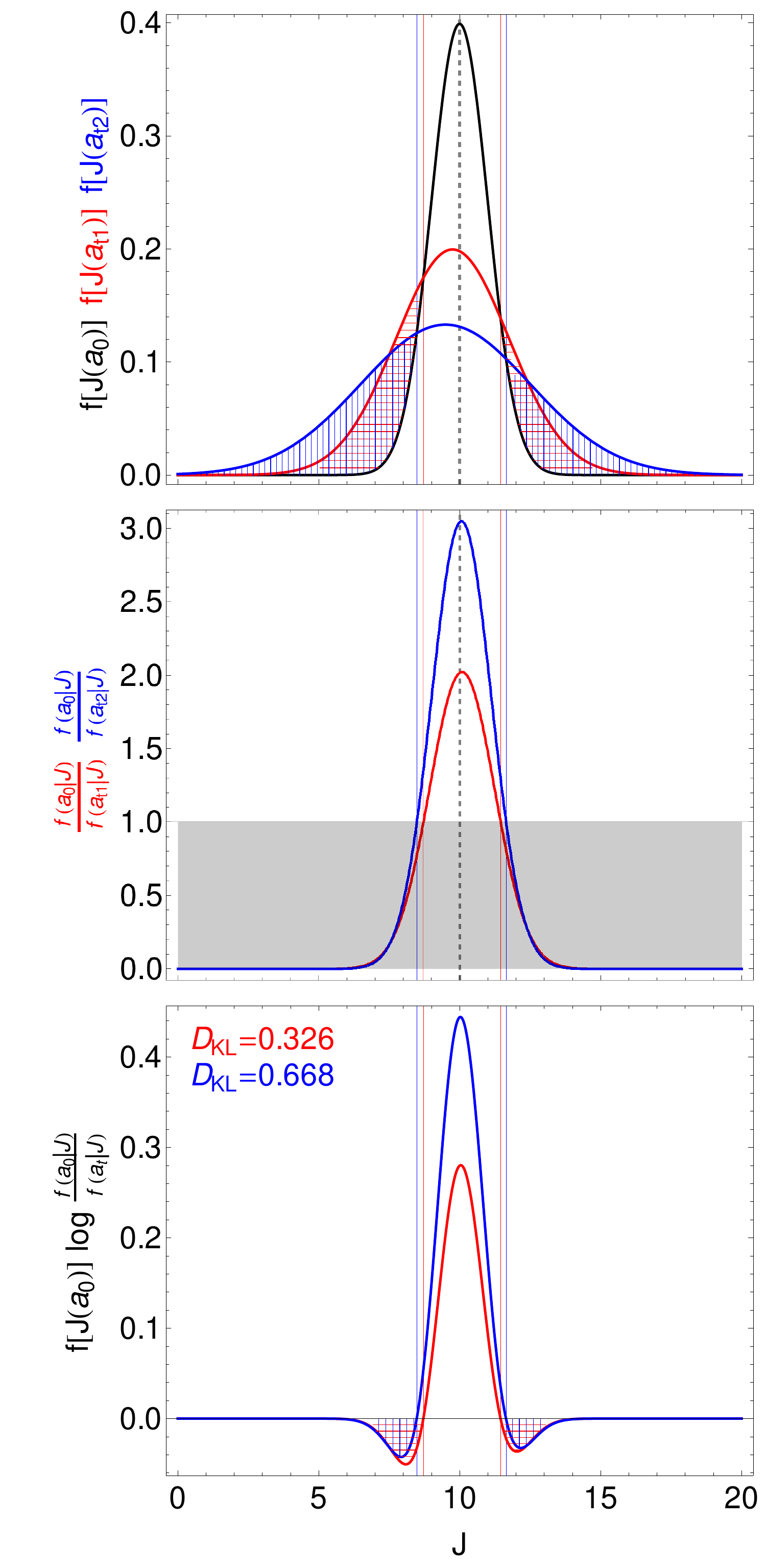}
\caption{Graphical illustration of steps in computing the KLD between action distributions for different values of the potential parameters.}
\label{fig:1dexample}
\end{figure}

\acknowledgements
The authors are grateful to Maarten Breddels (Kapteyn) for providing the {\sc HyperQuadTree} code to sample parameter space. It is a pleasure to thank
  Jo Bovy (IAS),
  Kathryn Johnston (Columbia),
  Hans-Walter Rix (MPIA), 
  Sanjib Sharma (Sydney), and Kyle Westfall (Kapteyn/Portsmouth)
for valuable conversations, and Eite Tiesinga and Wim Zwitser for technical support of the computing cluster at the Kapteyn Institute, which was used to explore parameter space. The authors also thank the anonymous referee for a careful review that led to a deeper understanding of the determination of uncertainties for this unusual estimator.
RES and AH gratefully acknowledge support from the European Research Council under ERC-Starting Grant GALACTICA-240271. DWH acknowledges support from the NSF (IIS-1124794) and NASA (NNX12AI50G).

\appendix
\section{Setting confidence intervals with the KLD: one-dimensional example}
\label{appx:KLDIllustration}

As discussed in the main text, we set confidence intervals by comparing the distribution of the actions with the best-fit parameters, $f_{\vect{a}_0}(\vect{J})$, with the distribution of the actions for some other trial parameters, $f_{\sub{\vect{a}}{trial}}(\vect{J})$. In this example we will consider a single stream described by a one-dimensional Gaussian in a single action variable $J$ (top panel of Figure \ref{fig:1dexample}). We further assume we have already maximized the KLD of this distribution with respect to some comparison distribution to find the best-fit parameters. (In more than one dimension we would maximize $\super{\sub{D}{KL}}{I}$ as defined in Equation \eqref{eq:dkl1}, but in a single dimension one cannot construct a product of marginals.) Therefore in the best-fit potential the Gaussian is tallest and thinnest (black line in top panel). For other values of the parameters the Gaussian's width will increase and its center can also shift slightly. We show distributions for two sets of trial parameters (red and blue lines in top panel), $\vect{a}_{t1}$ and $\vect{a}_{t2}$, with progressively less resemblance to the best-fit distribution. The red distribution is twice as wide as the best-fit and centered 0.25 to the left of the black one; the blue distribution is three times as wide and offset by 0.5. 

At any single value of $J$, the ratio of the trial and best-fit distributions is the ratio of their probabilities. This means that wherever $f_{\vect{a}_0}(\vect{J}) > f_{\sub{\vect{a}}{trial}}(\vect{J})$ (unshaded regions in the top panel), the best-fit parameters are a better fit than the trial parameters at that $J$, while when $f_{\sub{\vect{a}}{trial}}(\vect{J}) > f_{\vect{a}_0}(\vect{J})$ (red and blue hatched regions in the top panel), the trial parameters are a better fit. An alternative view is shown in the center panel of Figure \ref{fig:1dexample}, where the ratio of the two is plotted. The shaded region where the ratio is less than one shows for which $J$ the distribution with trial parameters is a better fit than the best-fit parameters (outside the vertical red or blue lines, respectively). At the $J$ where the best-fit parameters are most preferred over the trial parameters, the best fit is about twice as likely as our example $t1$ and about three times as likely as $t2$.

In order to get the expectation value of the probability ratio, i.e., the most likely value of the probability ratio for any $J$, we have to weight this ratio by the distribution of $J$ in the best-fit case. This is shown in the bottom panel of Figure \ref{fig:1dexample}. This quantity is the integrand in the KLD. Regions where the integrand is less than zero (red and blue hatched) denote $J$ where the best-fit is outdone by the trial parameters. However as we can see from the top panel, most of these values of $J$ are not very probable, so their contribution to the expectation value is small compared to the contribution from regions where 1) $J$ is probable \emph{and} 2) the best-fit is better than the trial parameters at describing the distribution. Integrating up this function gives the values of the KLD for the two trial parameters shown in the upper left-hand corner of the bottom plot. The red curve is closer to the best-fit distribution than the blue one (shown in the top panel of the figure), and so the red curve has a lower KLD than the blue one. We conclude that the $J$ are a bit less than half as likely (since $\log(2) \approx 0.69$) to be drawn from the blue distribution as from the black one, while the red curve is slightly more probable than that (about a 1 in 1.4 chance of producing the black curve).

\section{Estimating confidence intervals via the Hessian of the KLD}
\label{appx:KLDHessian}

In cases where computing $\sub{\super{D}{II}}{KL}$ over the entire parameter space is computationally expensive, the KLD also provides a shortcut to error estimates. A parameterized model distribution $\sub{f}{model}(\vect{a})$ can be fit to an empirical distribution $\sub{f}{emp}$ by \emph{minimizing} the KLD 
\begin{equation}
\sub{\super{D}{IIa}}{KL} = \int \sub{f}{model}(\vect{x} | \vect{a}) \log \frac{\sub{f}{model}(\vect{x} | \vect{a})}{\sub{f}{emp}(\vect{x})} d\vect{x},
\end{equation}
since the KLD approaches zero for identical distributions as discussed above. Furthermore when using this method to find the best-fit parameters of the model distribution, then the Hessian of $\sub{\super{D}{IIa}}{KL}$ at the best-fit point $\vect{a}_{0}$ is equal to the Fisher information matrix $\mathcal{I}$ \citep[][Section 2.6]{kullback1959}: 
\begin{equation}
 \left.\frac{\partial^2 \sub{\super{D}{IIa}}{KL}}{\partial a_i \partial a_j}\right|_{\vect{a_0}} = \mathcal{I}_{ij},
\end{equation}
whose inverse is the Cram\'er-Rao lower bound on the variance of the estimator. Therefore, in cases where one can ensure   a single maximum in parameter space (as opposed to several disconnected regions of high probability), and the region around the maximum is well-behaved enough to admit approximation by a Taylor series, the variance obtained by inverting the Fisher information can be used as an approximation of the error.

For our work, we can consider the ``empirical" distribution to be the distribution of the $\vect{J}$ for $\vect{a}_{0}$, and the ``model" distribution to be that of $\vect{J}$ with some other potential parameters $\sub{\vect{a}}{trial}$. In this case the minimum of the KLD is zero, and the Hessian of 
\begin{equation}
\sub{\super{D}{IIa}}{KL} = \int f_{\sub{\vect{a}}{trial}}(\vect{J}) \log \frac{f_{\sub{\vect{a}}{trial}}(\vect{J})}{f_{\sub{\vect{a}}{0}}(\vect{J})} d^3\vect{J},
\end{equation}
evaluated at the best-fit point, gives us the Fisher information for our best-fit parameters. Note that this is $\sub{\super{D}{II}}{KL}$ with the arguments reversed; for small $\Delta\vect{a}$ near the best-fit point the KLD is symmetric. 

\bibliography{KLClusters}

\end{document}